\documentclass[usenatbib]{mnras}

\usepackage{graphicx,ulem}
\usepackage{amssymb}
\usepackage{amsmath}
\usepackage{xspace}
\xspaceaddexceptions{]\}}
\usepackage{todonotes}
\usepackage{multirow,booktabs,bigdelim,mathtools}
\usepackage{eso-pic}

\newcommand{\sect}[1]{Sect.~\ref{#1}\xspace}
\newcommand{\sects}[1]{Sects.~\ref{#1}\xspace}
\newcommand{\app}[1]{Appendix~\ref{#1}\xspace}
\newcommand{\fig}[1]{Fig.~\ref{#1}\xspace}
\newcommand{\figs}[1]{Figs.~\ref{#1}\xspace}
\newcommand{\eq}[1]{Eq.~(\ref{#1})\xspace}

\newcommand{\tab}[1]{Table~\ref{#1}\xspace}

\newcommand{\nn}{\nonumber\\}

\usepackage{lineno}

\newcommand{\lneqb}{\begin{linenomath*}}
\newcommand{\lneqe}{\end{linenomath*}}

\newcommand{\lcdm}{\ensuremath{\Lambda}CDM\xspace}
\newcommand{\wcdm}{\ensuremath{w}CDM\xspace}
\newcommand{\mpp}{\ensuremath{3\!\times\!2}pt\xspace}
\newcommand{\cosmosis}{{\sc CosmoSIS}\xspace}

\newcommand{\datavect}{\ensuremath{\vec{d}}\xspace}
\newcommand{\datat}{\ensuremath{d}\xspace}

\newcommand{\datavecm}{\ensuremath{\hat{\vec{d}}}\xspace}
\newcommand{\datam}{\ensuremath{\hat{d}}\xspace}
\newcommand{\datavecblm}{\ensuremath{\hat{\vec{d}}_{\rm bl}}\xspace}

\newcommand{\cov}{\ensuremath{C}\xspace}
\newcommand{\covdat}{\ensuremath{C_{\datavect}}\xspace}
\newcommand{\params}{\ensuremath{\Theta}\xspace}
\newcommand{\pref}{\ensuremath{\Theta_{\rm ref}}\xspace}
\newcommand{\pobs}{\ensuremath{\Theta_{\rm obs}}\xspace}

\newcommand{\pblind}{\ensuremath{\Theta_{\rm bl}}\xspace}
\newcommand{\punblind}{\ensuremath{\Theta_{\rm unbl}}\xspace}
\newcommand{\likeli}{\ensuremath{{\cal L}}\xspace}
\newcommand{\modspace}{\ensuremath{{\cal M}}\xspace}
\newcommand{\dspace}{\ensuremath{{\cal D}}\xspace}
\newcommand{\prior}{\ensuremath{\operatorname{\pi}}\xspace}
\newcommand{\prej}{\ensuremath{\operatorname{Prej}}\xspace}
\newcommand{\prejspace}{\ensuremath{{\cal M}_{\prej}}\xspace}
\newcommand{\dpriorspace}{\ensuremath{{\cal D}_{\prior}}\xspace}
\newcommand{\dprejspace}{\ensuremath{{\cal D}_{\prej}}\xspace}
\newcommand{\ythree}{Y3-3\ensuremath{\times}2pt\xspace}
\newcommand{\yone}{Y1-3\ensuremath{\times}2pt\xspace}
\DeclareMathOperator*{\argmax}{argmax} 
\newcommand{\dchisq}{\ensuremath{\Delta\chi^2}\xspace}
\newcommand{\chisq}{\ensuremath{\chi^2}\xspace}

\renewcommand{\vec}[1]{{\mathbf{#1}}}

\begin{document}

\title{Blinding multiprobe cosmological experiments}


\author[Muir et al]{
\parbox{\textwidth}{
\Large
J.~Muir,$^{1,2,\star}$
G.~M.~Bernstein,$^{3}$
D.~Huterer,$^{2}$
F.~Elsner,$^{4,5}$
E.~Krause,$^{6}$
A.~Roodman,$^{1,7}$
S.~Allam,$^{8}$
J.~Annis,$^{8}$
S.~Avila,$^{9}$
K.~Bechtol,$^{10,11}$
E.~Bertin,$^{12,13}$
D.~Brooks,$^{4}$
E.~Buckley-Geer,$^{8}$
D.~L.~Burke,$^{1,7}$
A.~Carnero~Rosell,$^{14,15}$
M.~Carrasco~Kind,$^{16,17}$
J.~Carretero,$^{18}$
R.~Cawthon,$^{11}$
M.~Costanzi,$^{19,20}$
L.~N.~da Costa,$^{15,21}$
J.~De~Vicente,$^{14}$
S.~Desai,$^{22}$
J.~P.~Dietrich,$^{23,24}$
P.~Doel,$^{4}$
T.~F.~Eifler,$^{6,25}$
S.~Everett,$^{26}$
P.~Fosalba,$^{27,28}$
J.~Frieman,$^{8,29}$
J.~Garc\'ia-Bellido,$^{9}$
D.~W.~Gerdes,$^{30,2}$
D.~Gruen,$^{31,1,7}$
R.~A.~Gruendl,$^{16,17}$
J.~Gschwend,$^{15,21}$
W.~G.~Hartley,$^{4,32}$
D.~L.~Hollowood,$^{26}$
D.~J.~James,$^{33}$
M.~Jarvis,$^{3}$
K.~Kuehn,$^{34,35}$
N.~Kuropatkin,$^{8}$
O.~Lahav,$^{4}$
M.~March,$^{3}$
J.~L.~Marshall,$^{36}$
P.~Melchior,$^{37}$
F.~Menanteau,$^{16,17}$
R.~Miquel,$^{38,18}$
R.~L.~C.~Ogando,$^{15,21}$
A.~Palmese,$^{8,29}$
F.~Paz-Chinch\'{o}n,$^{16,17}$
A.~A.~Plazas,$^{37}$
A.~K.~Romer,$^{39}$
E.~Sanchez,$^{14}$
V.~Scarpine,$^{8}$
M.~Schubnell,$^{2}$
S.~Serrano,$^{27,28}$
I.~Sevilla-Noarbe,$^{14}$
M.~Smith,$^{40}$
E.~Suchyta,$^{41}$
G.~Tarle,$^{2}$
D.~Thomas,$^{42}$
M.~A.~Troxel,$^{43}$
A.~R.~Walker,$^{44}$
J.~Weller,$^{23,45,46}$
W.~Wester,$^{8}$
and J.~Zuntz$^{47}$
\begin{center} (DES Collaboration) \end{center}
}
\vspace{0.4cm}
\\
\parbox{\textwidth}{Affiliations are listed at the end of the paper.}
\\
\parbox{\textwidth}{$^{\star}$Email: jlmuir@stanford.edu}
}

\date{\today}

\label{firstpage}
\pagerange{\pageref{firstpage}--\pageref{lastpage}}


\AddToShipoutPictureBG*{%
  \AtPageUpperLeft{%
    \hspace{0.75\paperwidth}%
    \raisebox{-3.5\baselineskip}{%
      \makebox[0pt][l]{\textnormal{DES-2019-0427}}
}}}%

\AddToShipoutPictureBG*{%
  \AtPageUpperLeft{%
    \hspace{0.75\paperwidth}%
    \raisebox{-4.5\baselineskip}{%
      \makebox[0pt][l]{\textnormal{FERMILAB-PUB-19-563-AE}}
}}}%

\maketitle

\begin{abstract}
  The goal of blinding is to hide an experiment's critical results --- here the inferred cosmological parameters --- until all decisions affecting its analysis have been finalized. This is especially important in the current era of precision cosmology, when the results of any new experiment are closely scrutinized for consistency or tension with previous results. In analyses that combine multiple observational probes, like the combination of galaxy clustering and weak lensing in the Dark Energy Survey (DES), it is challenging to blind the results while retaining the ability to check for (in)consistency between different parts of the data.  We propose a simple new blinding transformation, which works by modifying the summary statistics that are input to parameter estimation, such as two-point correlation functions.  The transformation shifts the measured statistics to new values that are consistent with (blindly) shifted cosmological parameters while preserving internal (in)consistency. We apply the blinding transformation to simulated data for the projected DES Year 3 galaxy clustering and weak lensing analysis, demonstrating that practical blinding is achieved without significant perturbation of internal-consistency checks, as measured here by degradation of the $\chi^2$ between the data and best-fitting model.  Our blinding method's performance is expected to improve as experiments
    evolve to higher precision and accuracy.
\end{abstract}

\begin{keywords}
  cosmology: observations -- cosmology: large-scale structure of Universe --
  methods: data analysis -- methods: statistical -- methods: numerical
\end{keywords}

\section{Introduction}
\label{sec:intro}

The practice of blinding against human bias in data analysis is standard
in many areas of science.  The goal is to prevent the scientists from
biasing their analysis toward results that are theoretically expected or, more
generally, deemed  to be likely or correct. In experimental
particle physics strategies for  blinding are manyfold and have been honed since
their earliest application decades ago~\citep{Arisaka:1992xy}.  Blinding
strategies in particle physics include hiding the signal region, offsetting
parameters in the analysis by a hidden constant, and adding or removing events
from the analysis (for a review, see \citet{Klein:2005di}).

Blinding started to be applied to astrophysics and cosmology only relatively
recently. The first application  to cosmology was described in 
\citet{Conley:2006qb}, which reports on an analysis of magnitude-redshift data of Type Ia
supernovae (SNe Ia). In that study, the full analysis was performed with unknown offsets added to the   key cosmological parameters,  $\Omega_M$ and $\Omega_\Lambda$, until  unblinding revealed final parameter values. Many SNe Ia
analyses have adopted some variation of this blinding approach since
(e.g.\ \citet{Kowalski:2008ez,Suzuki:2011hu,Betoule:2014frx,Rubin:2015rza,Zhang:2017aqn,Abbott:2018wog}).
More recently, blinding has been regularly applied to analyses involving
strong gravitational lensing~\citep{Suyu:2012aa,Suyu:2016qxx},
as well as cosmological inferences from weak gravitational lensing
observations (e.g.\ \citet{Heymans:2012gg,vonderLinden:2012kh,Kuijken:2015vca,Blake:2015vea,Hildebrandt:2016iqg,Troxel:2017xyo,Hamana:2019etx}). 

For  cosmological analyses in general, direct application of blinding techniques from experimental particle physics is not feasible due to numerous  differences between cosmological observations and particle-physics experiments.
First, there is no clear division of the data space into a ``signal'' region that can be hidden vs a ``control'' region that can be used for all validation tests.
A second significant challenge arises from the fact that many cosmological inferences are now produced by combination of multiple ``probes,''
i.e.\ summary statistics of diverse forms of measurement of different classes of objects in the sky. For example, \citet{Abbott:2017wau} present a combined analysis of an observable vector \datavect composed of three  two-point correlation functions (2PCFs) measured from the first year of Dark Energy Survey (DES) data:  the angular correlation of galaxy positions via $w(\theta)$, the angular correlation of weak lensing shears via $\xi_{\pm}(\theta)$, and the cross-correlation between galaxy positions and shears via $\gamma_t(\theta)$. There is much degeneracy in how these diverse measurements contain cosmological information, which means that a simple blinding operation applied to one measured quantity can transform valid data into blinded data that are readily recognized as inconsistent with any viable cosmology.

The simplest form of blinding, which was used in the DES Year 1 galaxy clustering and weak lensing analysis, is to hide from users the values
  of the cosmological parameters that arise from the final inference,
  e.g. by shifting all values in any human-readable
results.  The risk of accidental revelation of the true parameter
estimates is high, however, if the blinding code is mistakenly
omitted.  The temptation for experimenters to peek at the true results is also high when the ``curtain'' is
so thin.  Furthermore, in this scenario, the blinding can potentially be compromised if anyone plots a
theoretical model on top of the measured summary
statistics.  It is therefore an advantage to apply a blinding
transformation at an earlier stage, when more steps are required to produce  unblinded results in a form that an experimenter
can recognize as conforming to their biases or not.

In this paper we propose a method for blinding multiprobe cosmological
analyses by altering the summary statistics which are used as input for
parameter estimation. The technique is very simply described by
\eq{eq:addbl}. This technique has the advantages of being
  applicable to observable vectors of arbitrary complexity while preserving
  internal consistency checks, and also of insuring that the inference
  code never even produces the true cosmological parameters until
  the collaboration agrees to unblind.
We are specifically developing and testing the
performance of this blinding scheme for the DES Year 3 combined probe analysis, but the ideas we present could in principle be
applied to any cosmological analysis.  Accordingly, we
 frame our 
discussion in terms of a generic experiment, beginning in \sect{sec:consider}
with a discussion of general considerations for blinding and how to assess
whether a blinding scheme can be successful. This is followed by \sect{sec:method}
where we introduce our summary-statistic-based blinding
transformation. Then,  \sect{sec:desmpp} describes the transformation's application within
the DES analysis pipeline as well as the results of
our tests of its performance for a simulated DES Year 3 galaxy clustering and
weak lensing analysis. We conclude in \sect{sec:conclusion}.
The data associated with the tests described below are available upon request.

\section{Prior and Prejudice: Considerations for blinding}
\label{sec:consider}

Broadly speaking, the goal of blinding is to change or hide the output of an analysis in a way that still allows experimenters to  perform validation checks on the analysis pipeline and data. Thus, in order to be effective, a given blinding scheme
must fulfill these requirements: it must be capable of altering
the analysis' output enough to overcome biases, and additionally must preserve the properties of
the data that are to be used in validation tests.
  Below we present three criteria that can guide the 
  determination of whether a given transformation of data can successfully
  blind an analysis. 

\subsection{Criterion I: concealing the true results}\label{sec:shiftenough}

Let us assume that the experiment produces a vector \datavecm of observed
quantities, and we wish to constrain the parameters \params of a
model $\datavect(\params)$ for these data.  The parameters can include astrophysical and
instrumental nuisance parameters as well as the cosmological parameters of
interest.  There will always be some prior probability,
$\prior{(\params)}$, that expresses the physical bounds of our model
(e.g. $\Omega_m>0$) and results of trusted previous experimentation.
In a Bayesian view, the purpose of the experiment is to produce a
likelihood function $\likeli(\datavecm | \params)$ which is combined with the
prior to produce a posterior measure of belief across the model space \modspace\ spanned by parameter vector $\params$,  
$P(\Theta |d) \propto \likeli(\datavecm | \params) \prior{(\params)}.$
One easily visualised variant of the prior is to have \modspace  be uniform
over some range of parameter values  and zero elsewhere, i.e.~\modspace\ is the parameter space encompassing all parameters \params considered feasible. 

The experimenters may additionally harbour prejudices about the
``correct'' values of the parameters; for instance that they should
agree with some theoretical  framework such as a flat Universe, or that
they should agree with some previous experiment that one is trying to
confirm.  We can express these 
prejudices with another (albeit, difficult to quantify) probability function $\prej(\params)$. It 
could for example be a uniform distribution over some region
$\prejspace \subset \modspace$.  Note that in this framing, one must make a decision
regarding previous  experiments' results: either we accept them as true
and place them in $\prior$; or we are using their comparison to our results as a test of our model, in which case
we must be wary of confirmation bias and should place them in \prej.

\begin{figure}
  \centering
\includegraphics[width=\linewidth]{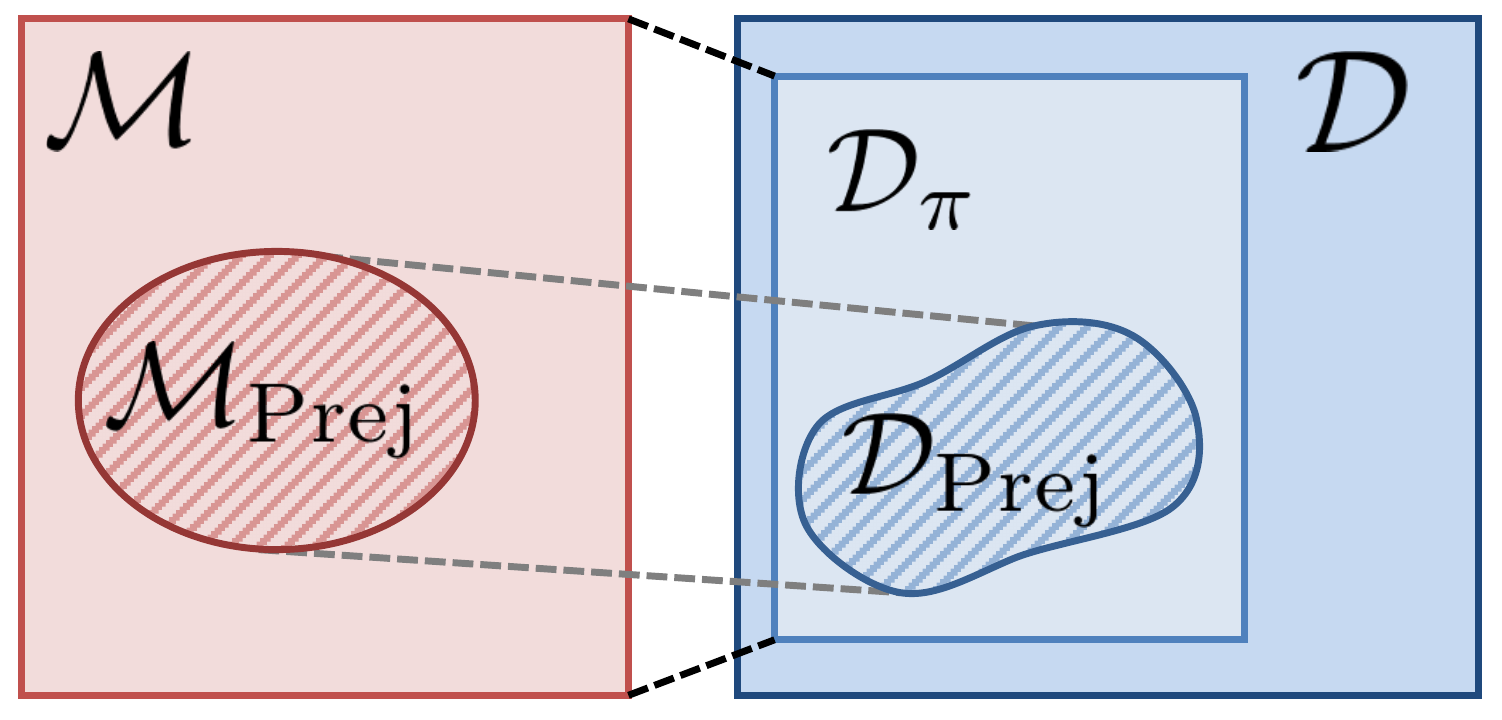}
  \caption{Cartoon of model and data spaces
    that we consider when thinking about how to blind an
    analysis, as discussed in \sect{sec:shiftenough}. \modspace\ is the space of all viable model parameter sets
    \params, which projects onto the observable-vector space
    $\dpriorspace\subset\dspace$, where \dspace\ is the space of all possible
    observable vectors. \prejspace\ is a region in parameter space associated with what we refer to as the ``prejudice'' distribution, describing 
    experimenters' preconceived expectations for where \params is likely to be.
    This subset of parameter space projects onto  $\dprejspace\subset\dpriorspace$.
    An effective blinding
    transformation must have the possibility of moving the observable
    vector \datavecm in or out of \dprejspace\ without moving it out
     of the prior space \dpriorspace. 
    }
  \label{fig:cartoon}
\end{figure}

The danger of experimenter bias arises when choices about the
analysis process are made, consciously or otherwise, on the basis of
whether the experiment's results conform to the prejudices,
i.e. whether $\params \in \prejspace.$
To confound the experimenter bias, a blinding procedure will apply a
transformation
\lneqb\begin{equation}
  \datavecm \rightarrow \datavecblm=B(\datavecm)
\end{equation}\lneqe
  to the data before
the experimenters perform analyses.  The first critical property of
$B$ is therefore that it must confound the experimenter's ability to know
whether the data are consistent or inconsistent with their prejudices.  For example, if we
take the maximum-posterior parameter values for blinded and
unblinded data
\lneqb\begin{equation}
\begin{aligned}
\punblind & = \argmax_{\params} \left\{P\left(\params | \datavecm\right)\right\} \\
\pblind & = \argmax_{\params} \left\{P\left(\params | B(\datavecm) \right)\right\},
\label{eq:maxpost}
\end{aligned}
\end{equation}\lneqe
then there must be a non-negligible chance that either
\lneqb\begin{equation}
\prej(\punblind)\gg \prej(\pblind)\quad
\textrm{or} \quad
\prej(\punblind)\ll \prej(\pblind).
\label{eq:prop1}
\end{equation}\lneqe
 A  graphical illustration is given in \fig{fig:cartoon}: If
we define \dprejspace\ as the region of data space
\dspace\ produced by parameter values within the 
prejudice region \prejspace, then
the blinding transformation must be able to move data into and out of
this region.  The experimenters should believe that this is possible,
but not know for certain whether it has happened.
Because we are dealing with human psychology and prejudices
  which may not be quantifiable, we
  usually cannot create a strict numerical requirement to satisfy this
  criterion.
  
\subsection{Criterion II: preserving the ability to check for errors}\label{sec:preserveval}

In addition to obscuring the true parameter output of an analysis, an
effective blinding scheme must still allow experimenters to examine the data \datavecm, before unblinding,
to uncover errors in their analysis procedure.  A validation test is one whose failure
indicates that data could not have been produced by any
allowed parameters $\params \in \modspace$. A  blinding transformation $B$ 
should not alter the conclusions of validation tests. 

There are a number of ways of stating this requirement. Sometimes the validation tests are expressed as some
projection of the data onto a ``null test'' $T(\datavecm)$ such that
\lneqb\begin{equation}\label{eq:nulltest}
T(\datavecm) = 0 \quad \forall \params \in \modspace.
\end{equation}\lneqe
Many kinds of validation tests fall into this paradigm. For example, if $T$ projects onto the B mode
(divergence-free component) of weak 
lensing, it should be zero within errors.
Another example is
  that a properly extinction-corrected galaxy survey should exhibit no
  statistically significant correlation between galaxy positions and
  star positions, so $T$ in this case would be the angular star-galaxy
  correlation function.
Or, $T$  can measure the
difference between observable vectors split by some property
presumed to be uncorrelated
with extra-galactic signals, such as seeing or the season when
  the data were collected.  Another very generic test is to
run the parameter inference on two subsets of the observable vector and
check that the results are consistent with common \params. Allowances must of course be made for the expected noise in the null
test output at fixed \params. Generally speaking,
a useful blinding transformation must yield
\lneqb\begin{equation}
T(\datavecblm)\sim 0
  \qquad\mbox{if and only if}\qquad
  T(\datavecm)\sim 0
\end{equation}\lneqe
where the $\sim$ sign implies consistency with zero within measurement errors.

More generally, $B(\datavecm)$  should map the allowed region \dpriorspace\ onto itself, and
likewise for its complement, the disallowed region $\tilde{\cal
  D}_{\prior}$.
Equivalently, the maximum-posterior values from
Equations~\ref{eq:maxpost} should obey
\lneqb\begin{equation}
\frac{ \likeli\left(\datavecblm | \pblind \right)}{
  \likeli\left(\datavecm_{\rm unbl}  | \punblind \right)} \approx 1.
\label{eq:prop2}
\end{equation}\lneqe
 In other words, blinding transformation should not significantly change the maximum likelihood in the parameter  space.  A transformation satisfying   this requirement will ensure that blinding will not alter experimenters' judgment  about  whether
there are flaws in the data.

\subsubsection{Model dependence of validation tests}\label{sec:modeldep}

It is important to note that defining validation tests requires one to make
implicit modeling choices, and defining a blinding procedure that preserves
the result of those tests can only produce shifts in parameter space that 
respects those choices. 
When constructing an analysis pipeline,
it is therefore important to carefully consider what measurements will be
considered results and which can be used as checks on the performance of the analysis pipeline. In other words, defining validation criteria and a blinding scheme that
preserves them requires one to specify the space of
models that are considered viable.

The only case in which a purely internal validation comparison can be
made without reference to a model is if the exact same observable is
measured twice [e.g. the amplitude of a particular cosmic microwave background (CMB) harmonic
measured at a particular frequency].  Any time two distinct quantities
enter the observable vector, a model is required to constrain their
joint distribution. 
For example, suppose we
compare the high- and low-redshift halves of a supernova Hubble
diagram.  If both halves are fit with a Lambda cold dark matter (\lcdm) model and the
data truly are from a \lcdm universe, then analysing the two
halves separately should produce consistent cosmology results, making this comparison a useful
validation test.  If, however, the universe is not described by 
\lcdm, then the high/low $z$ split can yield inconsistent
results even in the absence of processing errors. (Note that the original discovery of dark energy was effectively a demonstration that fitting supernovae data with $\Lambda=0$  produced this kind of mismatch.)

More subtly, a closer examination of the B mode null test described
  above (below \eq{eq:nulltest}) is another demonstration of how even
  nominally ``simple'' tests can be model-dependent. Though it is true that at leading-order 
  weak lensing in GR should have no divergence-free component, B modes can in
  fact be created by galaxy intrinsic alignments (IA) or modifications
  to GR. Thus, the test
  requiring measured B modes to be zero within errors can be more
  accurately rephrased as the requirement that the measured B modes be
  consistent with allowed IA  and gravity models.

\subsection{Criterion III: feasible implementation}\label{sec:whatdata}

Before we can begin determining the effectiveness of a blinding transformation $B$, we must first choose  what data will be transformed. As a concrete example, for imaging surveys like DES the data start as pixel
values, which are converted into catalogued galaxy fluxes, shapes, etc. Those catalogues are in turn converted into summary statistics such as the photometric redshift distributions $n(z)$ and tomographic weak lensing correlation functions
$\xi_{\pm}(\theta)$. The summary statistics are then finally used to obtain the parameter estimates $\hat\params$ themselves.

The simplest case is simply for $B$ to
operate on the $\hat\params$. As noted in Section~\ref{sec:intro},
this ``parameter shift'' method has been used frequently, but it has
the drawback of being a very thin cover over the truth.  It fails, for
example, if the experimenters become familiar with the relation
between the observables and the cosmological parameters and are
viewing how changes to the analysis impact the observables.  It can also be
difficult to employ this method while insuring that multiple probes
are consistent with a common model.  There is incentive to move the
blinding transformation to an earlier stage of the analysis, where the
scientists are less likely to be able to recognize whether their
prejudices have been confirmed.

Blinding by
alteration of the pixel data is probably impossible, apart from
substituting an entire set of simulated data for the real one.
Blinding at the catalogue level is possible in some cases, namely when a
change in a cosmological parameter maps directly into a change in some
catalogued galaxy property.  
For the DES Year 1 shear-only analyses
\citep{Troxel:2017xyo}, all
galaxy ellipticities (and hence all inferred weak lensing shears) were
scaled by an unknown multiplicative factor.  This is approximately
equivalent to a rescaling of $\sigma_8,$ though only in the linear regime.
The possibility of catalogue-level transformations becomes remote,
however, as we conduct
multiprobe experiments with many correlated 
summary statistics, and as multiple model parameters require
blinding. We have not been able to find a catalogue
transformation that preserves the validity of the data for the
DES combined galaxy clustering and weak lensing analysis.
This has motivated us to develop an approach to blinding
which relies on a transformation of the summary statistics, described in more detail below.

Recently \citet{Sellentin:2019stv} proposed a likelihood-level blinding
via modifications of the full covariance matrix of the
  observable vector. It will be of interest to see if this
  alternative blinding method robustly satisfies criterion II from above --- preserving
  the results of all validation tests --- given that the
  \citet{Sellentin:2019stv} alteration to the covariance matrix is
  dependent upon the observed data.

\section{Proposed method: Blinding by modifying summary statistics}
\label{sec:method}
Here we propose a method for consistently blinding cosmological analyses by transforming the summary statistics used as input for parameter estimation. Because parameter estimation is done by comparing measured summary statistics to theoretical predictions, the software infrastructure for an experiment will naturally include tools for computing model predictions at various points in parameter space. Our blinding transformation makes use of those tools to translate shifts in parameter space to changes in the summary statistics. 

The blinding transformation works as follows. Let $\datam_i$ be element $i$ of a measured observable vector, and let
$\datat_i(\params)$ be the theoretically computed (noiseless)
value of that same element for model parameters \params.  We choose a known reference
model $\pref$ and a blinding shift $\Delta\params$ in the
cosmological parameters.
The blinding operation is then a simple modification of each element
$\datam_i$ of the observable vector,
\lneqb\begin{equation}
  \begin{aligned}
B(\datam_i) & = \datam_i + f^{\rm(add)}_i, \\
f^{\rm (add)}_i & = \datat_i\left(\pref + \Delta\params\right) - \datat_i\left(\pref\right).
  \end{aligned}
  \label{eq:addbl}
\end{equation}\lneqe

If the expected noise level  on \datavecm  does not vary much across the
parameter shift $\Delta\params,$ then it is true by construction that $B$
will map data generated at $\pobs$ into viable data for
$\pblind=\pobs + \Delta\params$ if the truth (\pobs)
is sufficiently close to the reference cosmology (\pref).
However, because \pobs is not known (in fact, this whole exercise is
designed to obscure it!) and because the observable vector generally is not
actually linear in parameters, it is not guaranteed that this blinding
transformation will satisfy the necessary criteria for successful
blinding. Its application to a given observable vector and parameter space thus
requires numerical validation.\footnote{For some observables it
    might be possible to blind using a multiplicative transformation,
    multiplying the observable vector entries $i$ by  $f^{\rm (mult)}_i  =
    \datat_i\left(\pref + \Delta\params\right) /
    \datat_i\left(\pref\right).$
Our tests show, however, that this would rescale the noise in the observable vector as well as the signal, and would lead to unpredictable behaviour if any components $\datat_i$ are close to zero. Thus, in most cases \eq{eq:addbl}'s additive transformation will be preferable.}
 
\subsection{Procedure for blinding at the level of summary statistics}
\label{sec:blprocedure}
An overview of the procedure for summary-statistic blinding is as follows: 
\begin{enumerate}
\item Choose a reference cosmology (and nuisance parameters)
  $\pref$ in the middle of the range of models considered
  feasible truths.
\item Select a (blind) shift $\Delta\params$ from a distribution
  broader than the preconceptions
  causing the confirmation bias.  For example if there is a
  theoretical prejudice that the dark energy equation of state parameter is $w=-1$, then $\Delta w$ should be capable
  of shifts four to five times the experiment's forecasted uncertainty in $w$.
\item For each summary statistic $\datat_i$ being used for cosmological
  inference, calculate the blinding factor $f_i$ using \eq{eq:addbl}. 
\item Hide the real data $\datam_i$ and give experimenters the
  shifted values $B(\datam_i)$ as per \eq{eq:addbl} with which to conduct all validation
  tests. 
\item After passing validation tests, unblind by using the original unblinded data $\datavecm$ to repeat the  inference of $\params.$
\end{enumerate}
\subsection{Evaluating performance}
\label{sec:needtotest}

The blinding technique that we propose is fully described by 
\eq{eq:addbl}. In practice, the implementation of our blinding algorithm depends on the choice of the summary statistic to which the blinding
factors $\vec{f}^{\rm (add)}$ are applied,  the reference parameters \pref, as well as
the probability distribution from which parameter shifts $\Delta\params$ are
drawn.
In order to test the performance for a given set of these choices, we must show the following:

\noindent (1) the blinding transformation is able generate shifts in
best-fitting model parameters large enough to overcome experimenters'
potential biases, as described in \sect{sec:shiftenough},
and 

\noindent (2) the blinded observable vector $\datavecm_{\rm bl}$ is
consistent with data that could be produced by \textit{some} set of
allowed model parameters, as discussed in \sect{sec:preserveval}.

We can test both of these requirements  by analysing simulated data according to the procedure below. 
\begin{itemize}
\item We choose a reasonable reference cosmology \pref as well as an
ensemble of ``true'' parameters associated with observed unblinded data
$\{\pobs^{(a)}\}$, where $a$
labels the realization. For these realizations, we also select an ensemble blinding shifts $\{\Delta\params^{(a)}\}$. For example, we may choose to use the reference
cosmology with the dark-energy equation of state $w_{\rm ref}=-1$, and in some
realization of our blinding test, we could choose $w_{\rm obs}^{(a)}=-0.875$
and $\Delta w^{(a)} = -0.031$.
\item For each realization, we  generate a noiseless synthetic measured observable vector by computing the theory prediction for the data at the input cosmology,
\lneqb\begin{equation}
  \datavecm^{(a)} = \datavect(\pobs^{(a)}).\label{eq:simdata}
\end{equation}\lneqe
In our representative example above, this corresponds to, for example, a predicted weak lensing shear 2PCF  evaluated at $w_{\rm obs}^{(a)}=-0.875$.
\item We then blind that data vector using \pref and $\Delta\params^{(a)}$ according
to the transformation \eq{eq:addbl} to obtain $\datavecblm^{(a)}$. That
  is, we  evaluate
\lneqb\begin{equation}
  \datavecblm^{(a)} = \datavect(\pobs^{(a)}) +
  \datavect(\pref + \Delta\params^{(a)})-\datavect(\pref).
  \label{eq:simdata2}
\end{equation}\lneqe
In our  example this corresponds to the sum of the 2PCF for $w_{\rm obs}^{(a)}=-0.875$ and one for $w_{\rm ref}+\Delta
w^{(a)}=-0.906$, minus the 2PCF for $w_{\rm ref}=-1$.  
\item By performing parameter estimation on $\datavecm^{(a)}$ we can find associated unblinded best-fitting parameters $\punblind^{(a)}$. (For noiseless data we expect $\punblind^{(a)}=\pobs^{(a)}$.) Likewise we can find the blinded best-fitting parameters  $\pblind^{(a)}$ by performing parameter estimation on $\datavecblm^{(a)}$. 
\end{itemize}

Studying the distribution of these best-fitting parameters for such a simulated
analyses allows us to assess the performance of the blinding
transformation. Point~(1) 
from above (that blinding must be able to produce large enough shifts in parameter space) is straightforward to check. Generally we expect that the input blinding shift will determine the shift in output best-fitting parameters, 
\lneqb\begin{equation}\label{eq:inputoutputshifts}
  \pblind - \punblind \approx \Delta\params.
\end{equation}\lneqe
If this is true, we can ensure the blinding transformation satisfies this requirement
simply by drawing $\Delta\params$ from a wide enough probability distribution in \modspace. By analysing an ensemble of simulated observable vectors we can explicitly check the extent to which \eq{eq:inputoutputshifts} holds. It is worth noting here that it does not matter if the relation in \eq{eq:inputoutputshifts} strictly holds: Blinding can still be effective as long as the output shifts $\pblind - \punblind$ and input shifts $\Delta\params$ span a comparable region of parameter space. 

For  point~(2), 
we propose using the quantity \dchisq, defined below, as a metric for testing whether the blinding transformation  defined in \eq{eq:addbl} preserves the results of validation tests.\footnote{Of course, if one is considering applying this blinding procedure to an analysis which will use specific null or consistency tests as unblinding criteria, one should additionally check that the results of those tests are preserved.} This statistic is defined as the difference between the minimum \chisq for the model's fit to blinded data and that of the fit to unblinded data:
\lneqb\begin{equation}\label{eq:dchisqgeneral}
  \dchisq \equiv \left\{-2\ln
\likeli\left( B(\datavecm)|\pblind \right)\right\} - \left\{- 2\ln
\likeli\left( \datavecm|\punblind \right)\right\}.
\end{equation}\lneqe
It is a measure of the extent to which blinding preserves the internal consistency of the different components of the observable vector. In other words, it quantifies how much of the error in the model's fit to blinded data comes from the blinding procedure itself. If we can confirm that \dchisq is sufficiently small for all realizations in our ensemble of simulated observable vectors, we can ensure that the blinding transformation satisfies \eq{eq:prop2} for the set of input parameters and blinding shifts considered.

\subsection{Leading-order performance}
\label{sec:lop}
In the context of the evaluation metric described above, a
  perfect blinding technique can shift the inferred
parameters by the bias-defeating amount with $\dchisq=0,$ i.e.\ no
change in the degree to which data obey the model.  It is clear that
\eq{eq:addbl} will be perfect if the data depend on the model in a
purely linear fashion.  Furthermore the parameter shift will be
simple, i.e. \eq{eq:inputoutputshifts} will attain equality.

Indeed in this linear regime with fixed covariance, blinding via \eq{eq:addbl} guarantees
the more general statement that
\begin{equation}
  \chi^2\left[B(\datavecm) | \pobs + \Delta\params\right] =
  \chi^2(\datavecm | \pobs),
\label{eq:chiinvar}
\end{equation}
which in turn will lead to the Bayesian evidence
\begin{equation}
  p(\datavecm) = \int d\params \, \likeli(\datavecm | \params)
  p(\params)
\end{equation}
being invariant under the blinding transformation to the extent that
the prior $p(\params)$ is invariant under shift by $\Delta\params$. 
In this limit of fixed multivariate Gaussian noise and linear
parameter shifts, the additive blinding yields data that are fully
indistinguishable from a shift in the truth cosmology.

The blinding technique of \citet{Sellentin:2019stv} differs in that
the measured observable vector \datavecm\ is left unchanged, but the covariance
matrix undergoes a data-dependent transformation
$C_d \rightarrow \breve C_d(\datavecm)$ 
such that the
blinded $\breve \chi^2(\datavecm | \params)$ is guaranteed to satisfy
\begin{equation}
  \breve \chi^2(\datavecm | \pref + \Delta\params) = \chi^2(\datavecm
  | \pref).
\end{equation}
This is not quite the same condition as demanding $\Delta\chi^2=0$ in
\eq{eq:dchisqgeneral}, since the latter operates at the best-fitting
$\params$ in both the blinded and unblinded cases.
The basic construction for $\breve C_d$ does not guarantee
that the analogous property to \eq{eq:chiinvar} will hold, i.e. we can
have
\begin{equation}
  \breve \chi^2(\datavecm | \params + \Delta\params) \ne \chi^2(\datavecm
  | \params) \quad \textrm{if} \quad \params \ne \pref,
\end{equation}
even in the linear regime. We can expect that the
\citet{Sellentin:2019stv} blinding is functionally perfect by the
definition of $\dchisq$ in \eq{eq:dchisqgeneral} being small,
for the true $\params$ sufficiently close to $\pref$.  Numerical
investigations will be of interest to see whether covariance-matrix
blinding yields sufficiently small $\dchisq$ over the full range
of $\params$ allowed by any particular experiment.

Moving beyond the linear regime,
in \app{app:lo} we calculate the parameter shifts and $\dchisq$
induced by a quadratic dependence of data on parameters.  The result
is that the shift in best-fitting parameters is no longer equal to $\Delta\params$, but rather acquires a leading term that is linear in the product $\Delta\params \times \left(\pobs-\pref\right). $  In other words we should not expect equality in  \eq{eq:inputoutputshifts}.  Keep in mind though, as we noted above, fulfilling this equality is not a goal of blinding.

The more important result is the scaling
\lneqb\begin{equation}
  \dchisq \sim \left|\Delta\params\right|^2  \left|\pobs-\pref\right|^2 / \left| \cov \right|,
  \label{eq:scaling}
\end{equation}\lneqe
where $\cov$ is the covariance describing measurement errors on the parameters.

We can therefore expect that the blinding will succeed (by having insignificant $\dchisq$) within some sufficiently small region around the origin in the plane of blinding shift $\Delta\params$ and ``truth shift'' $\pobs-\pref.$   This region must be large enough to allow the blinding shift to span \prejspace\ and for the truth shift to span \modspace.  Below, we will test whether this condition is met for the DES Year 3 analysis.

An important consequence of the scaling relations is that this blinding transformation will improve as our knowledge and experimental precision  evolve.  Let us assume that future experiments reduce measurement errors by a factor $\alpha<1$ ($\cov\rightarrow \alpha^2\cov$).  To the extent that statistical rather than systematic uncertainties limit constraining power, the blinding shift necessary to defeat prejudices will shrink in concert ($\Delta\params\rightarrow \alpha\Delta\params$) as will the width of the priors to future experiments  ($\pobs-\pref \rightarrow \alpha(\pobs-\pref)$).  Under \eq{eq:scaling} we see that $\dchisq\propto \alpha^2$ under this evolution.  Hence, to the extent that these trends hold, future experiments will, in fact, become easier to blind, given the same models and observables.\footnote{This scaling may not hold if, for example, tensions between previous experiments provoke prejudices which are large compared to the projected errors of the experiment. This would motivate blinding and truth shifts that shrink more slowly than measurement errors. If this shrinking is slow enough, preserving \chisq  could actually get more difficult. On the other hand, this would mean that we are in the fortunate position of having an experiment whose power is more than sufficient to resolve the tension. Generally though, we expect that future experiments will  restrict us to regions of parameter space where we can more accurately model the observables as linear in parameters, and thus where our blinding transformation's performance improves.}

\section{Application: Dark Energy Survey analysis of galaxy lensing and clustering}
\label{sec:desmpp}

Using the discussions in \sects{sec:consider} and~\ref{sec:method} as a guide,  we  will now test the summary-statistic blinding transformation for the DES Year 3 galaxy clustering and weak lensing combined analysis.  The goal of this exercise is twofold. First, it will serve as a concrete demonstration of how summary-statistic blinding can be implemented, and secondly,  we will validate the transformation's performance for use in the DES Year 3 analysis. 

DES is an imaging survey that, over the course of six years, has measured
galaxy positions and shapes in a 5000 ${\rm deg}^2$ footprint in the southern
sky. It is designed to use multiple observable probes to study the
properties of dark energy and to otherwise test the standard cosmological model, \lcdm. Those probes include  galaxy
clustering, weak lensing, supernovae, and galaxy clusters. Though the blinding transformation presented in this paper could be potentially useful for all of these cosmological observables, our focus in this paper is on the combined analysis of galaxy clustering and weak lensing shear.

For conciseness,  we will refer to this as the \mpp analysis, so named because in it three types of 2PCF are used as summary statistics. These 2PCFs are galaxy position-position, shear-shear, and position-shear angular two-point correlations measured from DES galaxy catalogues. Additionally, we will use Year 1 (Y1) to refer to the analysis of the first year of DES data, which covers a footprint of roughly 1300 $\rm{deg}^2$ and for which results are reported in \citet{Abbott:2017wau}, and Year 3 (Y3) to refer to the ongoing analysis of the first three years of data, which will cover the full 5000 $\rm{deg}^2$ footprint at a similar depth.  

 Below, we first briefly motivate the need for blinding in
 DES combined-probe analyses (\sect{sec:desblinding}) before describing the \mpp data and analysis pipeline in
 \sects{sec:desmpp_datavec} and~\ref{sec:desmpp_modeling}.  Then,
 \sect{sec:desvalidation} introduces our
 procedure for testing the performance of the 2PCF blinding transformation. \sects{sec:results_fid} and~\ref{sec:results_nuis} present the
 results. 

\subsection{The need for blinding in DES analyses}\label{sec:desblinding}

Two of the most powerful ways DES data can test the \lcdm model are via the constraints it can place on the dark energy equation-of-state parameter $w$ and on the amplitude of matter density fluctuations $\sigma_8$. The equation-of-state parameter describes the ratio of pressure and density of a fluid description of dark energy in \wcdm, an extension of \lcdm, which describes dark energy as a fluid. If dark energy behaves as a cosmological constant (as in \lcdm),  this parameter will take the value $w=-1$, while $w\neq-1$ means that the dark energy density evolves with time. 
The matter density fluctuation amplitude $\sigma_8$ is of interest because it and $\Omega_m$ are the mostly precisely constrained parameters for DES' measurements of structure in the $z\lesssim 1$ Universe. Comparing DES constraints in the $\sigma_8-\Omega_m$ plane to those extrapolated under \lcdm from Planck CMB measurements thus tests the ability of \lcdm to describe the evolution of the large-scale properties of the universe from early and late times. Given these tests, whether or not DES observables are consistent with the special value $w=-1$ in \wcdm parameter space or with Planck $\Omega_m$-$\sigma_8$ results in \lcdm are questions of particular interest for DES analyses.

The \yone \wcdm constraints are  consistent with $w=-1$, while the \lcdm results showed a suggestive offset from Planck in the  $\Omega_m$-$\sigma_8$ plane, with the DES preferring a lower value of $\sigma_8$.\footnote{In \citet{Abbott:2017wau} this offset is reported to be statistically insignificant, though exactly how such a tension should be quantified is a subject of some discussion~\citep{Handley:2019wlz}.} As the \ythree analysis will use three times the sky area, that increased statistical power will cause DES constraints to tighten, and the community will be closely watching how the results compare to $w=-1$ and to the Planck $\Omega_m$-$\sigma_8$ constraints.   Thus, the parameters that we are particularly interested in blinding are $w$ and $\sigma_8$.
In the tests below, for simplicity we focus on blinding transformations
  shifting only those two parameters. This is sufficient to blind the DES
  analysis, but we do note that one could easily and reasonably adjust the
  transformation to include shifts in any other parameter(s) --- $\Omega_m$,
  for example --- in order to further confound the DES-Planck comparison.

\def\arraystretch{1.2}      
\begin{table*}\label{tab:fidvalues}
  \centering
  \begin{tabular}{c | c c | c c | c c c }
    &  &&\multicolumn{2}{|c|}{Parameter estimation} &\multicolumn{3}{c}{Distribution for blinding tests }\\
    &Parameter & \pref & Search bounds & Prior & $\pref+\Delta\params$ & \pobs [fid.] & \pobs [Nuis.] \\ 

    \toprule
    \multirow{10}{*}{\shortstack{ cosmology\\  parameters} }
    
    &$\sigma_8$ & 0.834 &  [0.1, 2.0]  &flat &[0.734, 0.954]& Y3 Fisher & matches fid.\\
    &$w$ & -1.0 & [-2.0, -0.0]  &flat &[-1.5, 0.5]&Y3 Fisher& matches fid.\\
    &$\Omega_m$ & 0.295 & [0.1, 2.0]  &flat &-&Y3 Fisher& matches fid.\\
&$h$ & 0.6882 & [0.2,1.0] &flat&-&Y3 Fisher& matches fid.\\
    &$\Omega_b $& 0.0468 & [0.03, 0.07]  &flat &- & - &- \\
    &$n_s$& 0.9676 & [0.87, 1.07]  &flat&- & - &-\\
      &$\Omega_{\nu}h^2$& $6.155\times 10^{-4}$& [0.0006, 0.01]& flat &- & - & [0.0006,0.00322]\\
    &$N_{\text{ massive }\nu}$& 3 &-& fixed   &- & - &-\\
    &$N_{\text{ massless }\nu}$& 0.046&- & fixed  &- & - &- \\
      &$\tau$& 0.08 & -& fixed &- & - &-  \\

    \midrule
    \multirow{5}{*}{\shortstack{lens  \\galaxy bias} }
    
    &$b_1$ & 1.45 & [0.8, 2.5] &flat&-& Y3 Fisher & full prior \\
    &$b_2$ & 1.55 & [0.8, 2.5]&flat&-&Y3 Fisher & full prior \\
    &$b_3$ & 1.65 & [0.8, 2.5]&flat&-&Y3 Fisher & full prior \\
    &$b_4$ & 1.8 & [0.8, 2.5]&flat&-&Y3 Fisher & full prior \\
    &$b_5$ & 2.0 & [0.8, 2.5]&flat&-&Y3 Fisher & full prior \\
  \hline
  \multirow{1}{*}{\shortstack{shear  calib.}}
    &$m_{1\text{--}4}$& 0.012 & [-0.1, 0.1] & $\mathcal{N}(0.012, 0.023)$ &- & - &[-0.57,0.081]\\
      \hline 
      \multirow{3}{*}{\shortstack{intrinsic \\alignments}}
    &$A_{IA}$& 0.0 & [-5, 5]&flat&- & - &[0,1]\\
    &$\alpha_{IA}$& 0.0 & [-5, 5] &flat&- & - &[-4, 4]\\
      &$z_0^{(IA)}$& 0.62 &-& fixed &- & - &-   \\
      
      \midrule
      \multirow{4}{*}{\shortstack{source galaxy\\ photo-$z$ bias}}
      
    &$\Delta z_1^{\rm source}$& -0.002 & [-0.1,0.1]&  $\mathcal{N}(-0.001,0.016)$ &- & - &[-0.05, 0.046] \\
    &$\Delta z_2^{\rm source}$&-0.015   &[-0.1,0.1]&$\mathcal{N}(-0.019,0.013)$ &- & - &[-0.054, 0.024]  \\
    &$\Delta z_3^{\rm source}$& 0.007   &[-0.1,0.1]& $\mathcal{N}(0.009,0.013)$ &- & - &[-0.026, 0.04]  \\
    &$\Delta z_4^{\rm source}$& 0.018   &[-0.1,0.1]&$\mathcal{N}(-0.018,0.022)$ &- & - &[-0.048, 0.84]\\

      \midrule
      \multirow{5}{*}{\shortstack{lens galaxy\\ photo-$z$ bias}}
      
    &$\Delta z_1^{\rm lens}$& 0.002 & [-0.05, 0.05]& $\mathcal{N}(0.008,0.007)$&- & - &[-0.022, 0.026] \\
    &$\Delta z_2^{\rm lens}$& 0.001 &[-0.05, 0.05]&$\mathcal{N}(-0.005,0.007)$&- & - &[-0.020, 0.022]   \\
    &$\Delta z_3^{\rm lens}$& 0.003 &[-0.05, 0.05]&$\mathcal{N}(0.006,0.006)$&- & - &[-0.018, 0.024] \\
    &$\Delta z_4^{\rm lens}$& 0.0 &[-0.05, 0.05]& $\mathcal{N}(0.0,0.01)$&- & - &[-0.03, 0.03]  \\
    &$\Delta z_5^{\rm lens}$& 0.0 &[-0.05, 0.05]&$\mathcal{N}(0.0,0.01)$&- & - &[-0.03, 0.03]  \\
    \bottomrule
  \end{tabular}

  \caption{ Fiducial values and prior ranges for the DES \mpp analysis pipeline studied in this work along with the range from which parameters are drawn for the blinding tests performed in this study. Fiducial values and priors are chosen to be similar to the settings for the DES \yone analysis (with the exception of $w$, which uses a wider range here than in \citet{Abbott:2017wau}), and the fact that for these tests we search over $\sigma_8$ rather than $A_s$. The columns labelled ``Distribution for blinding tests'' display the range from which parameters are drawn to create the realizations of synthetic data and blinding factors used for the  blinding tests and are described in \sects{sec:paramdraws_fid} and~\ref{sec:paramdraws_fid}. The ``Y3 Fisher'' label in the \pobs column refers a multivariate Gaussian, distribution with the parameter covariance  estimated using a Fisher forecast, centred on the fiducial parameter values. For $\Omega_{\nu}h^2$, the upper bounds of its prior range and the upper bound of its ``Nuis.'' range correspond to $\sum m_{\nu}=0.93\text{ eV}$, and $0.3\text{ eV}$, respectively.  }
\end{table*}

\subsection{Observable vector and likelihood}\label{sec:desmpp_datavec}

\begin{figure}
  \centering
\includegraphics[width=\linewidth]{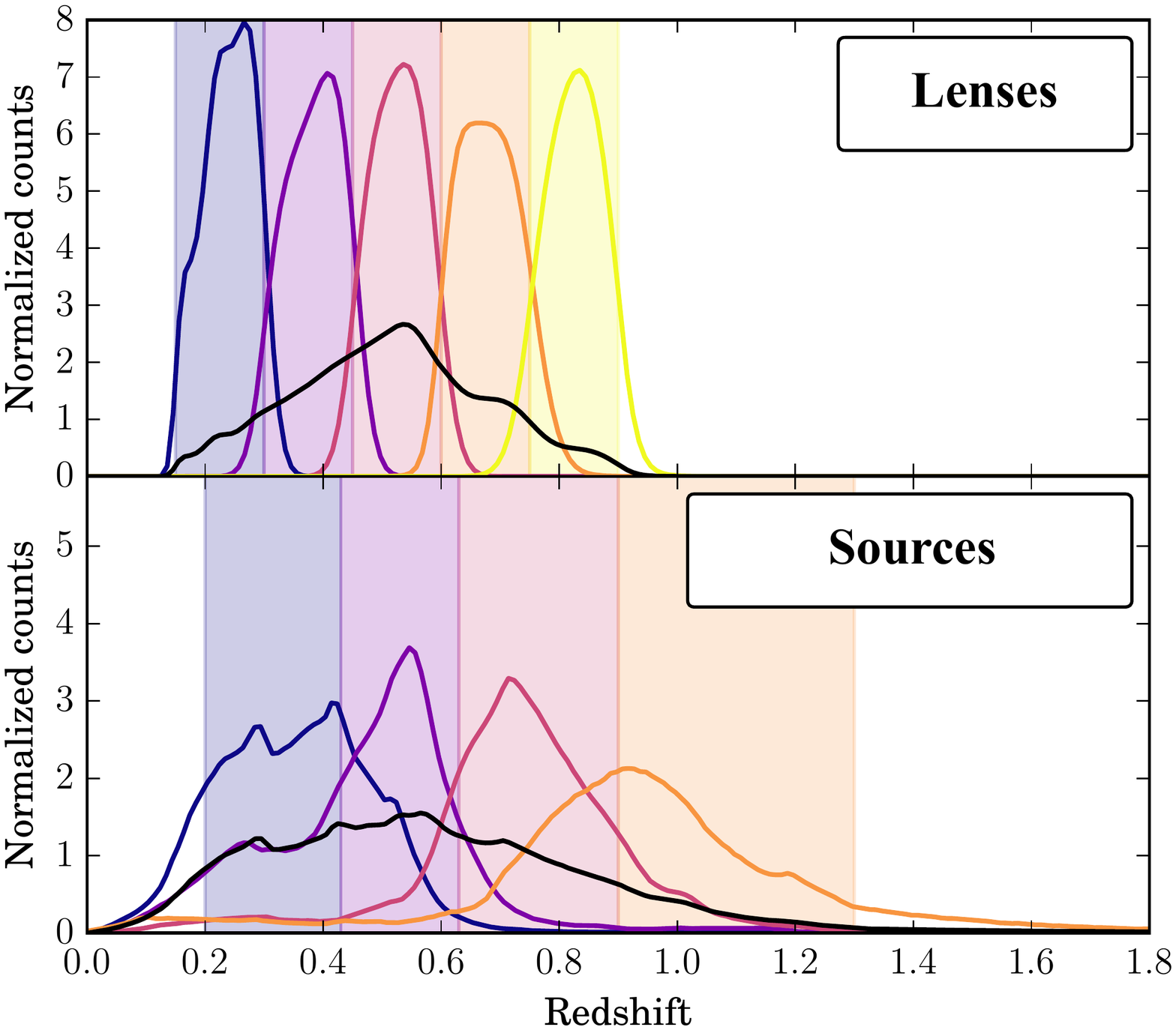}
\caption{
  The  $n(z)$ redshift distributions for lens and source galaxies in the DES \yone analysis from \citet{Abbott:2017wau}. The vertical coloured bands show the nominal redshift range of each bin, while the lines show the estimated true redshift distribution when galaxies are binned in photometric redshift. The black lines show the unbinned total distribution.  We adopt these same  redshift distributions for our Y3 blinding tests.
}
  \label{fig:dndz}
\end{figure}

Because the DES \ythree pipeline was not finalized when the blinding investigations presented in this paper were conducted, we approximate the Y3 observable vector and likelihood by using the Y1 modelling choices, prior ranges, and scale cuts. This section will briefly describe them, but we refer the reader to \citet{Abbott:2017wau}  and \citet{Krause:2017ekm} for a more detailed description of the associated measurements and calculations.

The DES \mpp analysis is based on observations of two populations of
galaxies. Positions are measured for a set of lens galaxies which have been
selected to have small photometric redshift (photo-$z$) errors and which have
been carefully checked for residual systematics. In the Y1 analysis, this
population consisted of 650,000 bright red-sequence galaxies which are
selected as part of the redMaGiC catalogue~\citep{Elvin-Poole:2017xsf}. Their
redshift distribution is shown in the upper panel of \fig{fig:dndz}.  Cosmic
shears are measured from a larger population of source galaxies. For the Y1
analysis the source galaxies included 26 million objects selected from the Y1
Gold catalogue~\citep{Drlica-Wagner:2017tkk} and their shapes are
measured as described in \citet{Zuntz:2017pso}.  The lens (source) galaxies are divided into
five (four) redshift bins, respectively; see \fig{fig:dndz}.

The \mpp observable vector consists of three kinds of 2PCF measured from the lens and source catalogues. The galaxy-galaxy correlations are measured as autocorrelations within each lens bin, producing a set of functions $w^i(\theta)$ for $i=1$--$5$. Shear-shear correlations are measured for all auto and cross-correlations of the source bins, producing functions $\xi_+^{ij}(\theta)$ and $\xi_-^{ij}(\theta)$ for $i=1$--$4$ and $1\leq j\leq i$.
The galaxy-shear cross-correlations are measured between all combinations of the five lens bins and four source bins, producing $\gamma_t^{ij}(\theta)$ for $i=1$--$5$ and $j=1$--$4$.
All of these 2PCF are measured for twenty logarithmically spaced angular bins
between 2.5 and 250 arcmin. Further scale cuts are applied in order to prevent modeling
uncertainties associated with non-linear structure formation, baryonic
physics, and other small-scale effects from biasing the final cosmological
results \citep{Krause:2017ekm}. The resulting \mpp observable vector has 457 entries.

The likelihood of the DES \mpp observable vector is modelled as a multivariate Gaussian.
Its covariance $\covdat$ has significant off-diagonal contributions, since many elements of the observable vector can share dependence on the realization of the mass and galaxy distributions in the survey volume (also known as sample variance).
We
adapt the covariance matrix that was previously analytically computed for the
Y1 analysis as described
in \citet{Krause:2017ekm} using  {\sc Cosmolike}~\citep{Krause:2016jvl}.  To approximate the Y3 covariance, we simply scale
the Y1 covariance  by a factor of $0.27=1350/5000$ to account for Y3's
increased survey area. This  survey-area scaling correctly modifies the
Gaussian parts of the covariance, but it does not properly scale the
non-Gaussian contributions~\citep{Joachimi:2007xd}. Thus, this is only a rough
approximation for the Y3 covariance, but it is sufficient for our testing purposes.
Though, in principle, the \mpp covariance depends on the model parameters, it has been shown~\citep{Eifler:2008gx} that the covariance's cosmology dependence can be neglected without significantly affecting parameter constraints. In the DES \yone analysis and in this work we do not vary the data covariance matrix when performing parameter searches.

\subsection{Modelling}\label{sec:desmpp_modeling}

The parameter estimation procedure for the DES \yone analysis, and therefore
also our simulated \ythree analysis, involves a search over 27 free parameters
for \wcdm. This includes seven cosmological parameters\footnote{Note that while
for the purposes of this blinding study we sample over $\sigma_8$ as an input
  model parameter, \citet{Abbott:2017wau} and other DES analyses typically
  sample over $A_s$ and measure $\sigma_8$ as a derived parameter.}
($\Omega_m$, $\sigma_8$, $w$, $h$, $\Omega_b$, $n_s$, and $\Omega_{\nu}h^2$)
and 20
nuisance parameters used to account for various systematic
uncertainties.  These nuisance parameters
include a constant linear galaxy bias $b_i$ for each of the five lens redshift
bins. Additional nuisance parameters are introduced to model the effects of
uncertainties in photo-$z$ estimation: for each lens and source bin, we
introduce a parameter $\Delta z_i$ that describes a redshift offset of that
bin's $n(z)$ distribution. To model shear calibration, we assign one
multiplicative shear calibration parameter $m_i$ per source galaxy
bin. Following the Y1 analysis, we impose tight Gaussian external priors on
all shear calibration and photo-$z$ nuisance parameters.  The last set of nuisance
parameters model how intrinsic (as opposed to lensing-induced) alignments
between galaxy shapes affect their observed 2PCF. We use a linear alignment
model with parameters $A_{IA}$, $\alpha_{IA}$, and $z_0^{(IA)}$. The
calculations we use to compute predictions for the \mpp observable vector given this
set of model parameters are described in \app{app:twoptcalcs} and in more detail in    \citet{Krause:2017ekm}.

 The fiducial values for all of these nuisance parameters, as well as cosmological parameters, are shown in \tab{tab:fidvalues}, as are the Gaussian priors used for the photo-$z$ shifts $\Delta z_i$ and shear calibrations $m_i$. During parameter estimation, the number of neutrinos $N_{\text{massive }\nu}$ and $N_{\text{massless }\nu}$ (chosen to sum to the standard model effective number of neutrinos $N_{\rm eff}$), the optical depth of the CMB $\tau$, and $z_0^{(IA)}$ are fixed, while the rest of the parameters shown in the table were varied with flat priors. 

\subsection{Evaluating performance for DES blinding}
\label{sec:desvalidation}

 \begin{figure}
   \centering
 \includegraphics[width=\linewidth]{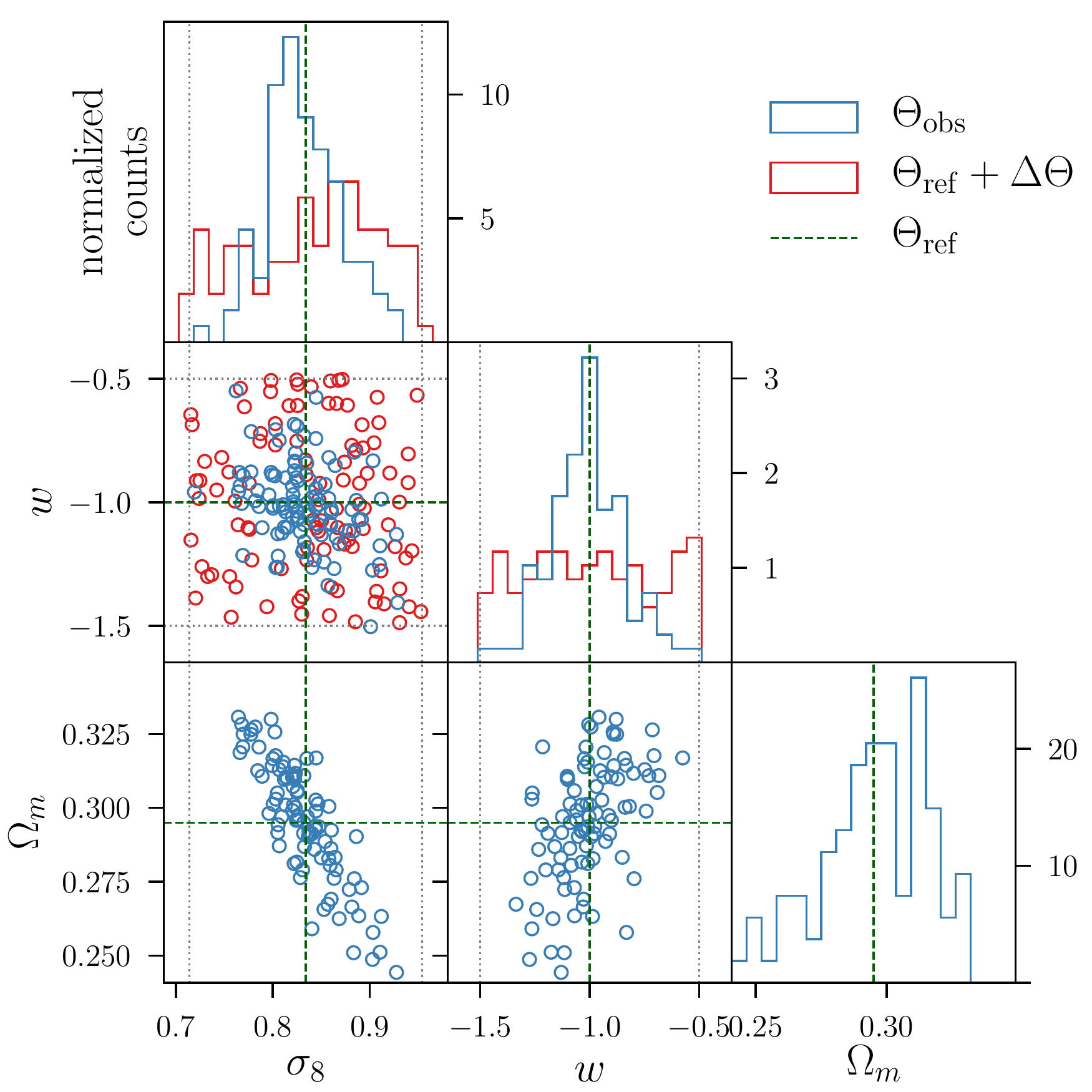}
 \caption{Values (off-diagonal panels) and the distribution (diagonal
     panels) of parameters $\sigma_8$, $w$, and $\Omega_m$ in the 100 realizations
     used for the fiducial test of our blinding procedure. The blue circles are the observed (that is
     true, unblinded) parameter values. The red circles are the shifted values
     used to blind the data. The dashed lines denote the reference model used
     for blinding. }
   \label{fig:parameterselection}
 \end{figure}

Our goal is to test the performance of 2PCF-based blinding for \ythree.   We do this by analysing an ensemble of 100 noiseless synthetic \mpp observable vectors.   For each realization, we draw $\Delta\params$ (which determines the blinding transformation) and \pobs (which determines the ``truth'' ) from probability distributions centred on  a reference cosmology \pref. That reference cosmology is fixed to the fiducial parameter values listed in \tab{tab:fidvalues}. The synthetic ``observed'' data \datavecm are then generated by computing a theory prediction for the \mpp observable vector at parameters \pobs, so
   \lneqb\begin{equation}
    \datam_i  = \datat_i(\pobs).
  \end{equation}\lneqe

  That data are then transformed according to \eq{eq:addbl} to produce a blinded observable vector,
  \lneqb\begin{equation}
     B(\datam_i) = \datam_i + \datat_i\left(\pref + \Delta\params\right) - \datat_i\left(\pref\right).
  \end{equation}\lneqe

We then search for parameters $\pblind$ and $\punblind$ that maximize the likelihood (minimize $\chi^2$)  for the blinded and unblinded data. The change \dchisq induced by applying the blinding transformation to the data will be our measure of success for the blinding transformation, since the blinded data should look compatible with some model in the parameter space. For select realizations, we additionally study the impact of blinding on the parameter estimation posteriors, as is shown in \fig{fig:chainplot_fid_nice}. The following subsections describe the technical details of this procedure.

\subsubsection{Parameter selection: fiducial test }
\label{sec:paramdraws_fid}

The $\Delta\params$ distribution is chosen to satisfy criterion I from \sect{sec:shiftenough} above, while \pobs will be drawn from a range that reasonably reflects potential offsets between the true \ythree cosmology and \pref.
We blind the two cosmological parameters that are at the greatest risk of experimenters' bias in the DES \mpp analysis: the amplitude of matter clustering $\sigma_8$ and the dark energy equation-of-state parameter $w$. We draw $\Delta\sigma_8$ from a flat distribution centred on zero with bounds $-0.12<\Delta\sigma_8<+0.12$ chosen to be roughly equal to the $3\sigma$ errors expected from \ythree.\footnote{For comparison, the \lcdm constraints on $\sigma_8$ reported in \citet{Abbott:2017wau} are $ 0.817^{+0.045}_{- 0.056}$.}  We draw $w$ from a flat distribution $-0.5<\Delta w<+0.5$,
chosen to span half of the flat prior being used for parameter estimation.  All other parameters have no input blinding shift.

For our fiducial test, we vary \pobs over a subset of the  27 \mpp \wcdm parameters: $\sigma_8$, $w$, $\Omega_m$, $h$, and the five lens galaxy bias parameters $b_i$ for $i\in \{1,\dots,5\}$.  It is drawn from a truncated multivariate Gaussian distribution in those parameters centred on \pref. The covariance of that distribution is obtained from Fisher forecast for our \ythree pipeline fixing all unvaried parameters. 
Any realization for which \pobs falls outside of the flat prior ranges in Table~\ref{tab:fidvalues} is discarded and redrawn.
\fig{fig:parameterselection} shows the resulting 100 realizations of
$\pref+\Delta\params$ 
and \pobs for a subset of parameters.

\subsubsection{\dchisq threshold}\label{sec:dchisqcutoff}

The blinded data should appear fully consistent with having been generated by some model within \modspace\ (here, \wcdm within the priors shown in \tab{tab:fidvalues}), if and only if the unblinded data are consistent with the true model.  Since \chisq is our measure of data-model consistency, this means that we would ideally like the blinding of the data to result in $\dchisq=0.$  A finite \dchisq is tolerable, however, if it is smaller than the expected statistical variation in the unblinded \chisq, i.e. \dchisq will not influence acceptance or rejection of the data.

We choose $\dchisq=30$ as a threshold for acceptable contributions from blinding to the error in the fit. This is within a few percent of the standard deviation $\sigma_{\chisq}\approx\sqrt{2\nu}$ 
for the number of degrees of freedom $\nu$ being
considered in our analyses.
For comparison, in the DES \yone analysis of \citet{Abbott:2017wau} an
unblinding criterion was that $\chi^2/\nu<1.4$. Our
$\dchisq=30$ threshold corresponds to $\dchisq/\nu=0.07$.

\begin{figure}
  \centering
  \includegraphics[width=\linewidth]{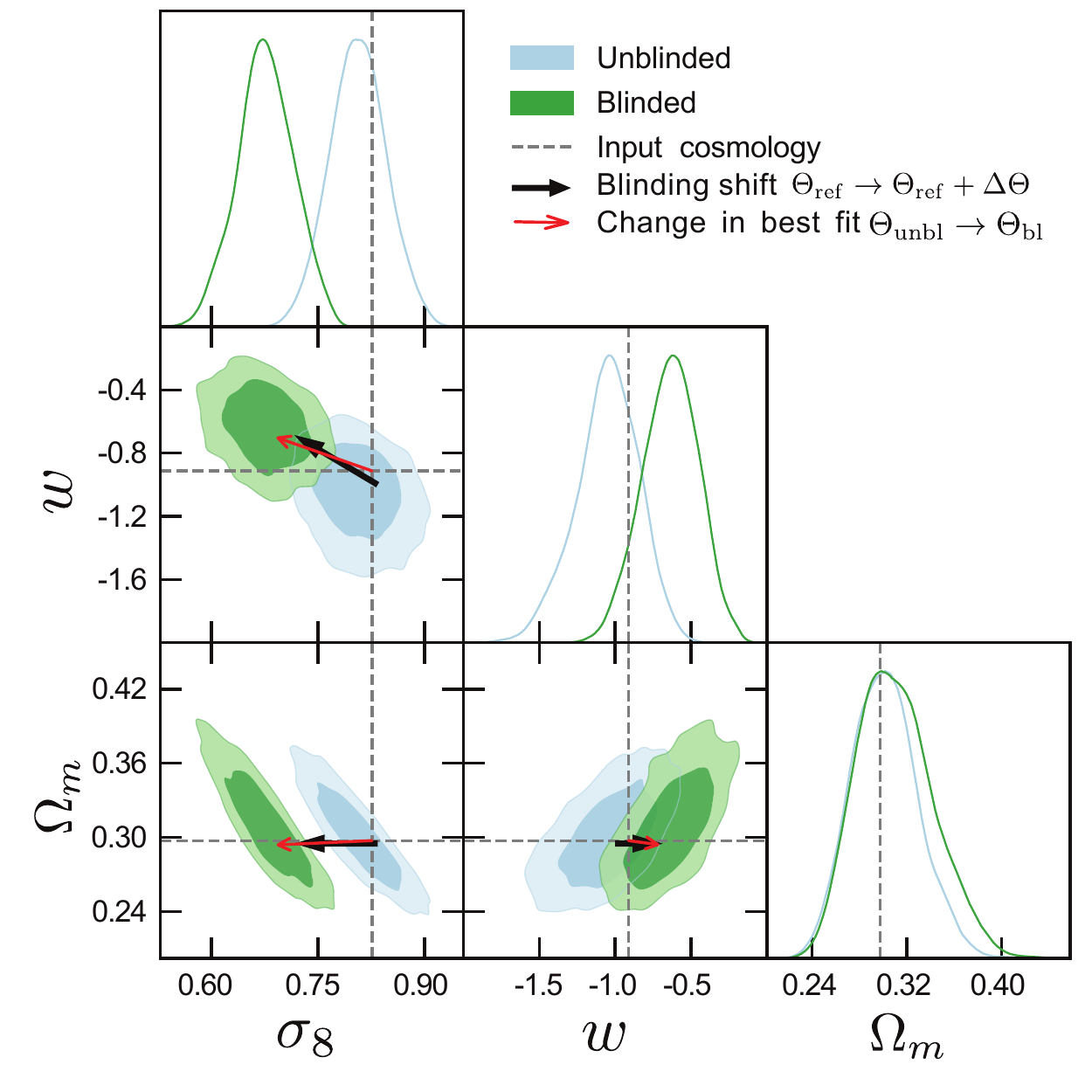}
\caption{The 68 and 95\% confidence intervals for blinded (green) and
  unblinded (blue) synthetic DES \ythree data. As an illustrative example,  results are shown for a
  realization of the fiducial blinding test with a large input
  blinding shift of $(\Delta\sigma_8 =-0.117,\Delta w=+0.314)$ and a
  low $\dchisq=1.86$. Dashed gray lines show the input parameters used
  to simulate the unblinded data $\pobs=(\sigma_8=0.826$, $w=-0.912$,
  $\Omega_m-0.29$). The black, thick arrow shows the points from \pref to $\pref+\Delta\params$ to show the input blinding shift, and the red, thin arrow points from $\punblind=\pobs$ to $\pblind$ to show the change in best-fitting parameters.  }
  \label{fig:chainplot_fid_nice}
\end{figure}
\subsubsection{Finding the maximum likelihood}\label{sec:findmaxlike}

To find the best-fitting parameters, we use the {\sc Maxlike} sampler in \cosmosis, which is a wrapper for the {\tt scipy.optimize.minimize} function
using the Nelder-Mead Simplex algorithm~\citep{Nelder:1965zz}.
This routine can fail to find the true maximum likelihood in high-dimensional
spaces, biasing $\chi^2$ high.  For our noise-free tests, we know $\chisq=0$ for the unblinded data, hence the measured \dchisq values are strict upper bounds on the true contributions of blinding to \chisq.
To more accurately characterise this bound, we re-run the \chisq minimisation search for all realizations with $\dchisq>30$ using  the {\sc Multinest} sampler\footnote{\tt ccpforge.cse.rl.ac.uk/gf/project/multinest/}~\citep{Feroz:2007kg,Feroz:2008xx,Feroz:2013hea} (as implemented in \cosmosis).   {\sc Multinest} is more computationally costly than Maxlike, but because it more thoroughly explores the parameter space it  is less susceptible to getting stuck in  local minima. We perform the {\sc Multinest} searches over the full 27-dimension \wcdm parameter space using the same low-resolution settings used for DES \yone exploratory studies\footnote{These {\sc Multinest} settings are: 250 live points, efficiency 0.8, tolerance 0.1.} and substitute the Maxlike results with those from  {\sc Multinest} results in cases where the minimum \chisq reported from  {\sc Multinest} is smaller.\footnote{Because {\sc Multinest} is designed to map  the posterior distribution rather than find the best-fitting point in parameter space, its accuracy will be limited by the density of samples in the high likelihood region. Based on {\sc Multinest} fits to unblinded observable vectors (where we know the true minimum is $\chisq=0$), we estimate that the {\sc Multinest} results tend to overestimate the minimum \chisq by $\sim 5$.}

\subsection{Results for fiducial test}
\label{sec:results_fid}

\fig{fig:chainplot_fid_nice} shows an example of how the parameter contours
shift in response to a large blinding shift of $(\Delta\sigma_8 =-0.117,\Delta w=+0.314)$. For this realization, the change in goodness of fit is small, $\dchisq=1.86$, and we can see that the location of the 68 and 95 percent confidence contours change in a way consistent with  the input blinding shift.

\begin{figure}
  \centering
\includegraphics[width=\linewidth]{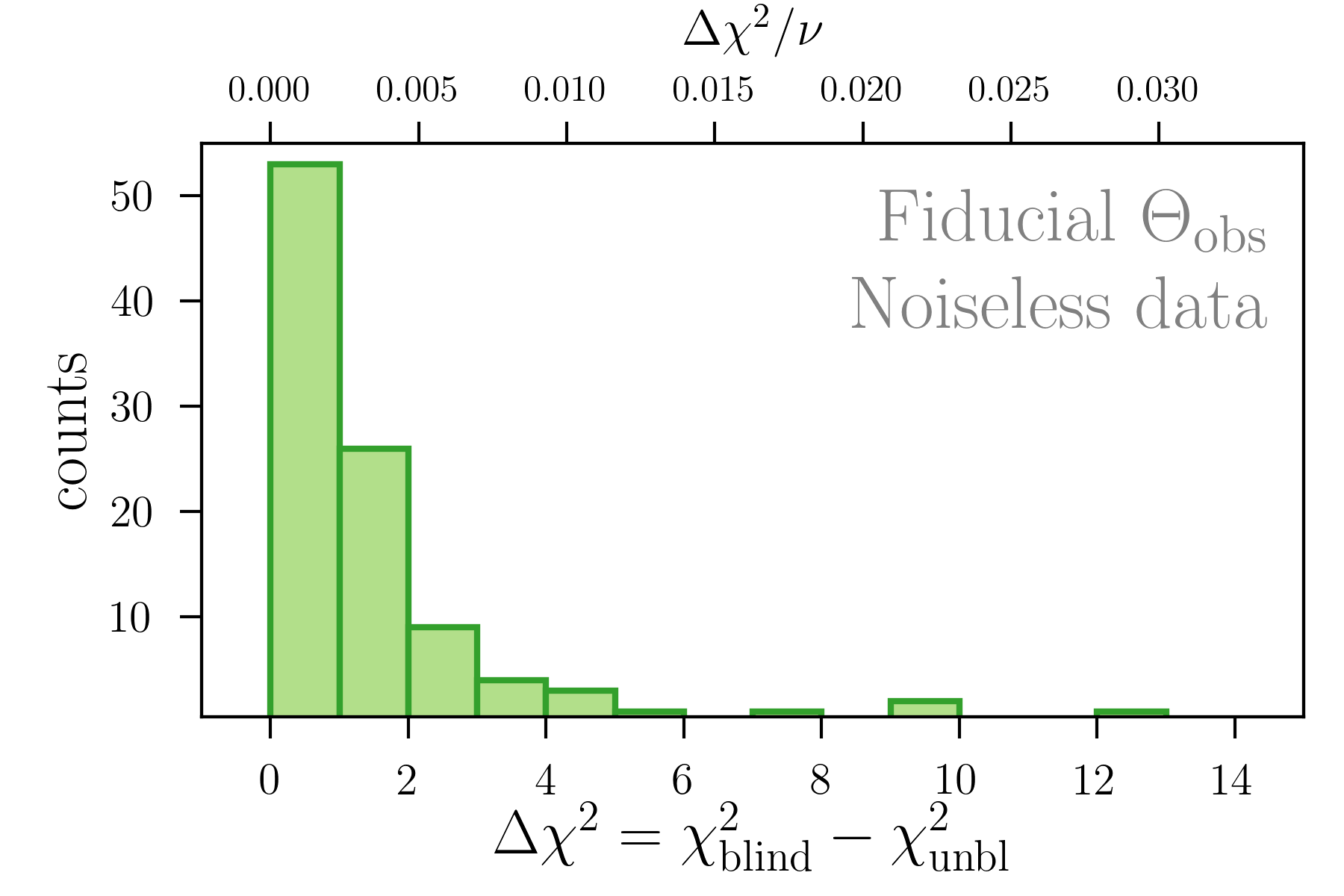}
\caption{Fiducial blinding test results for \dchisq. The top axis shows \dchisq in units of the degrees of freedom associated with the DES \ythree \wcdm analysis, $\nu=430$. 
}
  \label{fig:dchisqhist_fid}
\end{figure}

More quantitatively, \fig{fig:dchisqhist_fid} contains the primary results for our test of the 2PCF-based blinding for the DES \ythree analysis. It shows a histogram of the \dchisq values for the 100 blinded and unblinded pairs of synthetic observable vectors analysed. We see that in all realizations \dchisq is well below our $\dchisq=30$ cut-off. This means that
the blinded data are indistinguishable from unblinded data at different input parameters.%

 \begin{figure}
    \centering
\includegraphics[width=\linewidth]{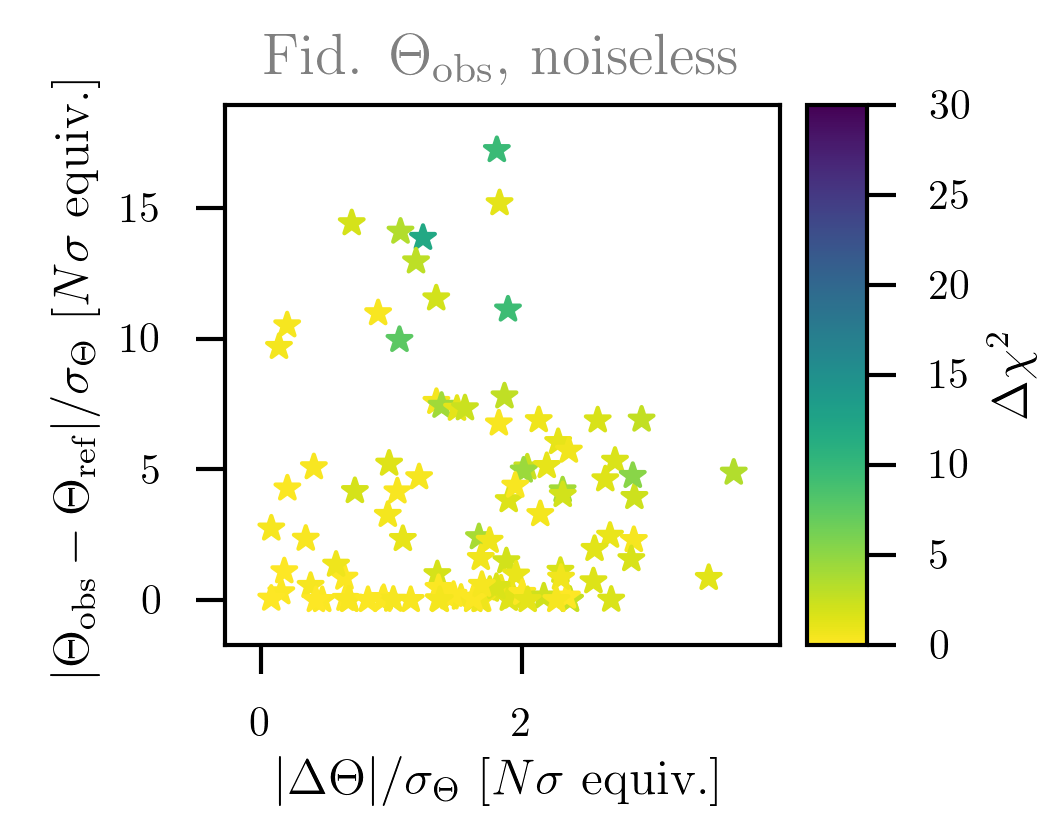}
\caption{Dependence of \dchisq on the magnitude of the truth shift $|\pobs-\pref|/\sigma_{\params}$ (vertical axis) and of the blinding shift $|\Delta\params|/\sigma_{\params}$ (horizontal axis) for the  observable vectors generated using the fiducial 100 realizations of true cosmology \pobs. These distances are evaluated in the two-dimensional $\sigma_8-w$ plane for the blinding shift and in the full 27-dimensional parameter space for the truth shift. In both cases the distance is scaled according to expected parameter uncertainties and are shown in units of the number of standard deviations for a one-dimensional Gaussian distribution with equivalent probability-to-exceed. The colours of the points represent the \dchisq value of those realizations.}
  \label{fig:dist-dist-dchisq_fid}
 \end{figure}

 In \sect{sec:lop} we suggested that the performance of the blinding transformation should worsen (higher \dchisq) when both the blinding shift $\Delta\params$ and the truth shift $\pobs-\pref$ become significantly non-zero.  \fig{fig:dist-dist-dchisq_fid} plots the \dchisq of each fiducial realization (as the colour) vs the size of the blinding shift and the truth shift.  To quantify the size of these shifts, 
we extract a parameter covariance \cov from a fiducial {\sc Multinest} chain and use it to compute distances in parameter space,
\lneqb\begin{align}
  |\pobs-\pref|/\sigma_\Theta &= \sqrt{(\pobs-\pref)^T\cov^{-1}(\pobs-\pref)},\\
  |\Delta\params|/\sigma_\Theta &= \sqrt{(\Delta\params)^T\cov^{-1}(\Delta\params)}.
\end{align}\lneqe
For the blinding shift, we compute this distance in the marginalized two-dimensional $\sigma_8-w$ plane (that is, we use a $2\times2$ \cov containing only the entries for those parameters). We evaluate the truth shift in the full 27-dimensional parameter space. For ease of interpretation,
these distances are then converted into an equivalent deviation for a one-dimensional Gaussian normal distribution by equating the probability-to-exceed value: we solve for $N$ such that
\lneqb\begin{equation}
  P(\chisq>N^2, \nu=1) = P(\chisq>\left(|\Delta\params|/\sigma_\Theta\right), \nu=2)
  \label{eq:pte}
\end{equation}\lneqe
and likewise for the truth shift (but with $\nu=27$ on the right).
For example, a truth shift of 
  $ |\pobs-\pref|/\sigma_{\params}=\sqrt{30}\simeq 5.5$ that corresponds to 
$\chisq=30$ in 27-dimensional parameter space has the same probability-to-exceed as a $\chisq=1$ signal in a one-dimensional Gaussian, so we would plot this as a ``$1\sigma$'' blinding shift in the metric of the parameter covariance matrix.

\fig{fig:dist-dist-dchisq_fid} confirms the behaviour derived in \sect{sec:lop} that the larger \dchisq values appear only when both $\pref+\Delta\params$
and \pobs move significantly away from \pref under the metric of the experimental posterior. 
This trend is non-monotonic because the performance of blinding
depends somewhat on the direction in parameter space of the vector
$\pobs-\pref$ in addition to its magnitude.  Furthermore there is
noise in our \dchisq evaluation from imperfect optimisation.

\begin{figure*}
  \centering
\includegraphics[width=\linewidth]{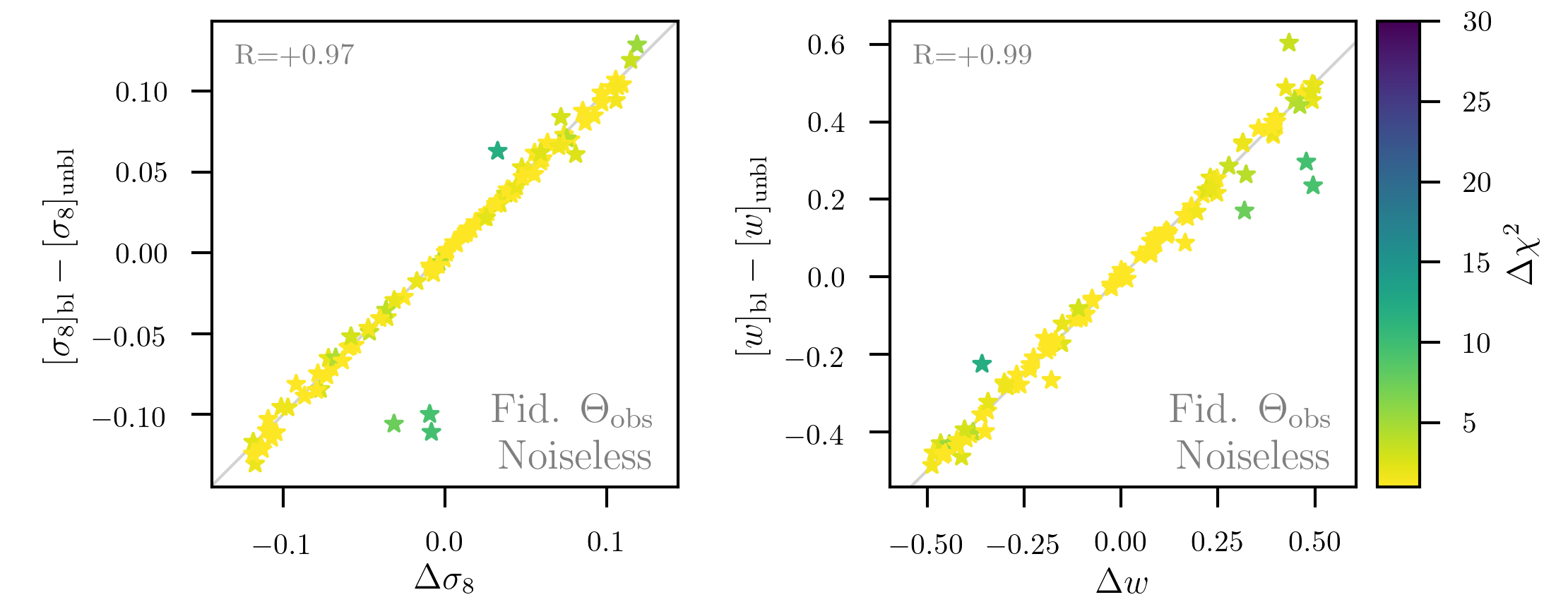}
\caption{Relationship between the input blinding
  shifts (horizontal axes) in $\sigma_8$ (left-hand panel) and $w$ (right-hand panel) and the output shifts in best
  fit values of those same parameters.  The colour scale shows \dchisq and the
  diagonal grey line shows the case where the output shifts are exactly equal
  to the input shifts. Both plots show data from the same 100 realizations for
  noiseless simulated data.}
  \label{fig:inoutshifts_fid}
\end{figure*}

\fig{fig:inoutshifts_fid} shows the relationship between the input blinding shifts in  $\sigma_8$ and  $w$ and the resulting shift in their best-fitting values. 
We generally find the behaviour expected from the leading-order analysis, which is for $\pblind-\punblind$  to be roughly equal to the parameter shift $\Delta\params$ 
used to generate the blinding factors, although with some deviation that grows roughly linearly with $\Delta\params$.  The scatter is expected because unplotted parameters, as well as the truth shift $\pobs-\pref$, also influence the deviation.
The fact that the range of the points along the veritical axes of  \fig{fig:inoutshifts_fid} is comparable to that along the horizontal axis does confirm that this blinding transformation is capable of altering the experiment's results enough to overcome experimenters' potential biases. 
Having satisfied all of the desired  criteria described in \sect{sec:consider}, we consider the blinding transformation of \eq{eq:addbl} to be successful for this fiducial analysis.

\fig{fig:inoutshifts_fid} can give us additional insight into the performance of our blinding transformation. 
In that Figure, the colour of each point corresponds to the \dchisq value for that realization.
 We note that the strong outlier points are also the points with largest \dchisq.  
 This is the expected behaviour. We saw in \fig{fig:dist-dist-dchisq_fid} that these high-\dchisq realizations have large truth shifts, and we expect that larger truth shifts should induce larger differences in $\pblind-\pobs$,
 mediated by the non-linear portions of the data model as indicated by \eq{tildetheta}. Those differences are expected to occur for all parameters, not just those we choose to blind. 
We confirm from our Maxlike results that, for such outlier realizations of
\pobs, blinding induces large changes in the best-fitting values of several
nominally unblinded parameters.

To further investigate this, we additionally
ran {\sc Multinest} chains for those realizations. We found that in two of these
realizations the 68 or 95 percent confidence intervals of posterior are pushed
into the hard-prior boundary for some of the galaxy bias parameters. Such
behaviour could be problematic, since checking that posterior distributions
have not hit prior bounds is a standard part of data validation, and the
blinding could trigger a false alarm for this. This issue not necessarily
prohibitive: Since we know that this kind of unpredictable shift is occurring
for realizations with large truth offsets $\pobs-\pref$, one could imagine
workarounds in the blinding procedure to account for, or  protect against, this possibility.  For example, the collaboration could ask a single member to confirm whether the collision with the prior bound is in fact due to the blinding shift or, as discussed in \sect{sec:conclusion} below, one could make use of several distinct blinding shifts defined at different \pref values.

\subsubsection{Impact of noise}
 
As an additional check, we perform this  analysis on a version of these same 100 synthetic observable vectors with Gaussian noise added using a Cholesky decomposition of the covariance \covdat. Results from this test are shown in \app{app:noisy}. Compared to the noiseless results shown above, there is slightly more scatter in the input-vs-output parameter shift relationship, and in the relationship between \dchisq and the various input parameters. These differences are expected and do not change the conclusions that the blinding transformation is effective.


\subsection{Follow-up test: varying nuisance parameters}
\label{sec:results_nuis}

As a stronger test of 2PCF-based blinding for the DES \ythree analysis, we analyse  a second set of synthetic observable vectors
for which more  parameters of \pobs are allowed to deviate from
\pref. This is a more rigorous test because it allows for
  larger differences between the
truth model and the reference model; recall that we noted in
  \sect{sec:lop} that we expect the performance of the blinding method to degrade as the magnitude of the difference $\pobs-\pref$ increases.
We use the same 100 blinding factors  as in the fiducial test. For the 100 realizations of \pobs, the values of $\sigma_8$, $w$, $\Omega_m$, $h$ match those used in the fiducial test, but  we additionally vary  $\Omega_{\nu}h^2$, reselect values of galaxy bias $b_{1\text{--}5}$, and vary all remaining nuisance parameters over flat probability distributions with ranges shown in the rightmost column of \tab{tab:fidvalues}.  For the nuisance parameters with Gaussian priors in the \mpp analysis, the ranges are $\pm 3\sigma$ for that prior.  The galaxy bias parameters are drawn from their full prior range, while the ranges for neutrino mass and intrinsic alignment parameters are chosen by eye based on the DES \yone posteriors in \citet{Abbott:2017wau}.  In the text and plots below we will refer to results from this test as the ``nuisance'' test (as opposed to fiducial test).

\begin{figure}
  \centering
\includegraphics[width=\linewidth]{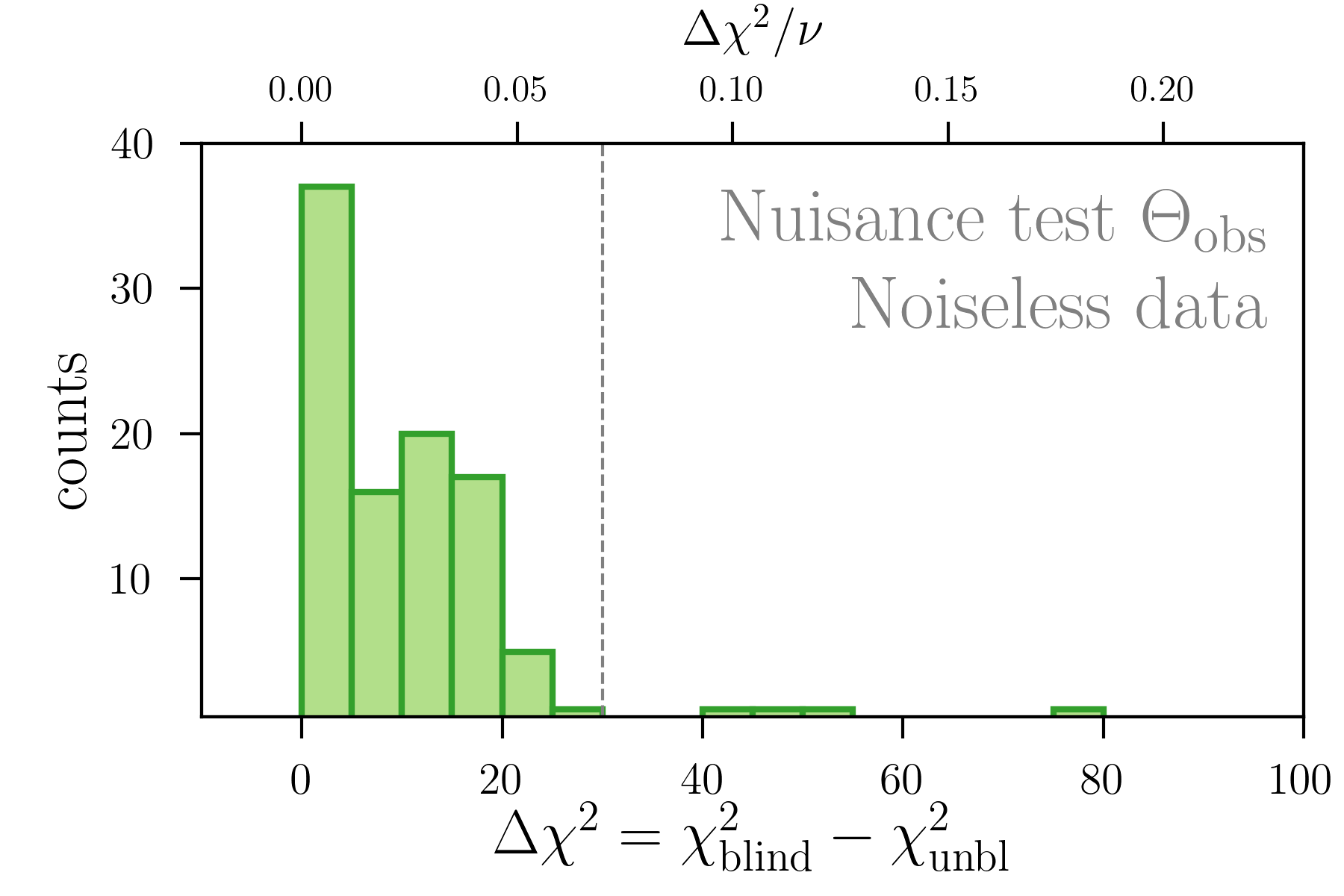}
\caption{Same as \fig{fig:dchisqhist_fid}, but for the ``nuisance test'' ensemble of unblinded observable vectors: histogram of \dchisq values.  The vertical dashed line shows our $\dchisq=30$ threshold, which is now exceeded by 4 of the 100 trials. }
  \label{fig:dchisqhist_nuis}
\end{figure}

After searching for best-fitting parameters\footnote{When we run {\sc Multinest} for select realizations of the nuisance test, we fit over the likelihood rather than the posterior. We do this because for many realizations the prior is very small at the true cosmology \pobs, causing significant differences between the maxima of the posterior and likelihood. This procedure is specific to the synthetic data study presented here where we are purposefully drawing \pobs from a wide range of nuisance parameter values, and where we are trying to minimize \chisq rather than maximize the posterior. For the analysis of real data, nuisance parameter priors should reflect our knowledge of what values they are likely to take, and one should maximize the posterior.} on the pairs of blinded and unblinded observable vectors of the nuisance test, we can study the same results discussed in \sect{sec:results_fid}. \fig{fig:dchisqhist_nuis} shows a histogram of the \dchisq values. Although the majority of realizations remain below  $\dchisq<30$, the distribution is broader than in the fiducial test and four realizations exceed that threshold. 

  \begin{figure}
    \centering
    \includegraphics[width=\linewidth]{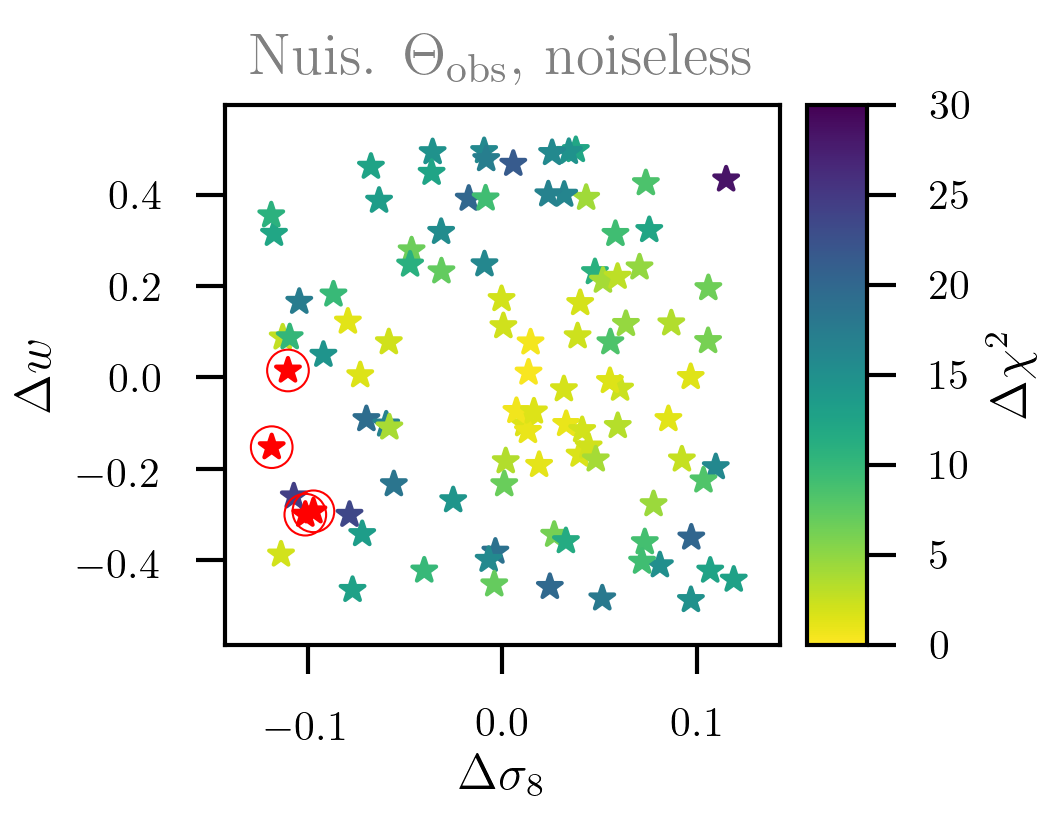}
\caption{For the ``nuisance test''
  ensemble, we plot the \dchisq induced by the blinding process as the colour of each point, vs the blinding shifts applied to the two cosmological parameters of greatest interest.  Points with $\dchisq>30$ are shown in red and are circled. }
  \label{fig:chisq_vs_blshift_nuis}
  \end{figure}

  \begin{figure}
    \centering
\includegraphics[width=\linewidth]{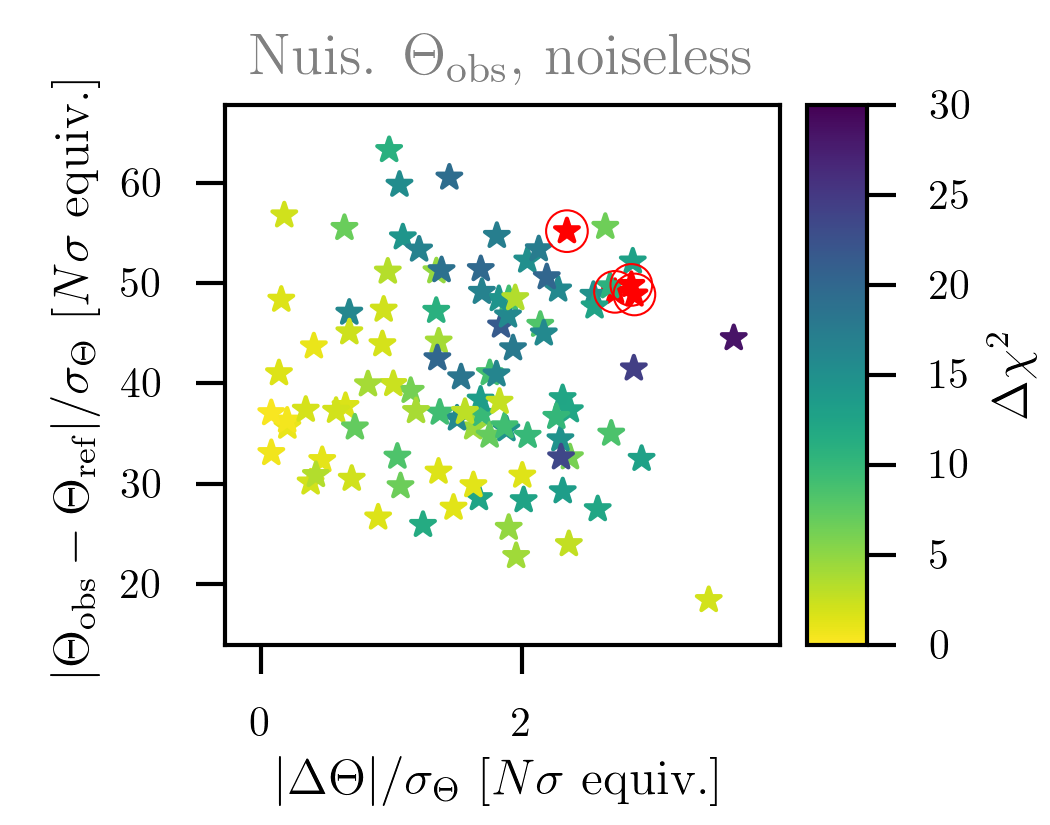}
\caption{Same as \fig{fig:dist-dist-dchisq_fid}, but for the ``nuisance test'' ensemble of unblinded observable vectors: dependence of \dchisq  on the parameter-space distances associated with $\pobs-\pref$ and $\Delta\params$. Circled points shown in red are realizations with $\dchisq>30$. }
  \label{fig:dist-dist-dchisq_nuis}
  \end{figure}

 \begin{figure*}
  \centering
\includegraphics[width=\linewidth]{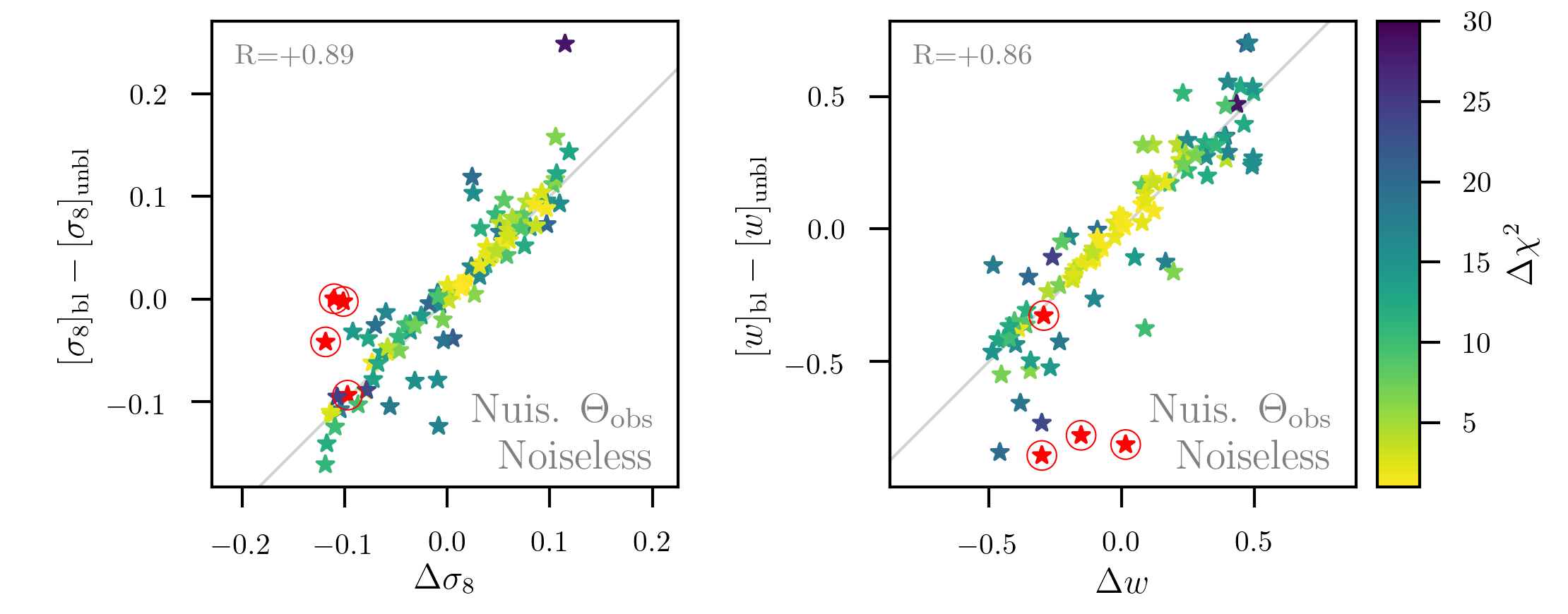}
\caption{Same as \fig{fig:inoutshifts_fid} but for the ``nuisance test'' ensemble of unblinded observable vectors: the relationship between input blinding shifts
  $\Delta\params$ and output shifts in best-fitting parameters $\pblind-\punblind$. Realizations with $\dchisq>30$ are shown with circled red
  points. The grey lines correspond to equal input and output shifts.}
  \label{fig:inoutshifts_nuis}
 \end{figure*}

  To understand the nature of the failures, we examine how \dchisq depends on the blinding shifts and truth shifts of each realization.  
  \fig{fig:chisq_vs_blshift_nuis} plots the distribution of \dchisq values for the nuisance test in the space of blinding shifts applied to $\sigma_8$ and $w$.  Here it is apparent that larger blinding shifts, particularly those that reduce $\sigma_8$, are associated with higher \dchisq.  A more revealing picture is given in
  \fig{fig:dist-dist-dchisq_nuis}, where it is clear that the unacceptable \dchisq trials are those with larger blinding shifts \emph{and} extremely large distances between \pobs and \pref (under the posterior parameter metric).  Note that change of scale on the $y$ axis from \fig{fig:dist-dist-dchisq_fid}.

In interpreting these results, we emphasize that this is by design a conservative test of the blinding transformation. 
The fact that the true cosmologies
\pobs are drawn from a flat probability distribution in a large number of nuisance
parameters means that the distances shown on the vertical axis of \fig{fig:dist-dist-dchisq_nuis} are very large relative to projected \ythree uncertainties. In other words, all of the realizations studied for the nuisance test have truth parameters that are collectively highly unlikely under the priors that have been assigned to them\footnote{The Gaussian nuisance prior probabilities $\pi$ evaluated at input truth values \pobs fall in the range $10^{-40}<\pi(\pobs)<10^{-10}$ for all nuisance test realizations.}, and even more unlikely under the expected posterior probability.  To put this in perspective, note that a more realistic simulation of the blinding transformation's performance could be done by drawing \pobs from the DES \yone posterior probability distribution, which might be considered an appropriate prior for Y3. The resulting ensemble of $\pobs-\pref$ separations would be significantly smaller than those used here, we would expect better performance in our \dchisq test. However, especially given the limited number (100) of realizations, sampling \pobs from the Y1 posterior would result in few realizations with large $\pobs-\pref$ which might test the limits of this approach to blinding. In contrast, our nuisance test's flat probability distributions and resulting extremely large truth shifts let us probe regions of parameter space where the blinding transformation breaks down.  We then check that those regions are highly unlikely for the application we are considering. We conclude that, although the nuisance test has failed blinding Criterion 2 of \sect{sec:preserveval} in 4\% of our trials, this only occurs for realizations with true parameters that are much farther from our reference parameters than they would realistically be for DES \ythree.

\fig{fig:inoutshifts_nuis} shows how blinding shifts relate to output shifts in the best-fitting parameters for the nuisance test. Compared to similar results from the fiducial test (\fig{fig:inoutshifts_fid}) there is more scatter in this relationship and the high-\dchisq points depart further from the trends. These behaviours are  consistent with what we expect from expanding the \pobs space.  There is some curvature present in these relations at low $w$, which suggests that the model is extending to regions where the quadratic approximation assumed in \app{app:lo} is inadequate.  Nonetheless we still confirm that the output parameter shifts are large enough to defeat observer prejudice. 

\section{Discussion and Conclusions}
\label{sec:conclusion}

This paper
presents and demonstrates
an effective blinding strategy for  multiprobe cosmological
analyses at the summary-statistic level, with the principal requirement that such a strategy allows for
robust  validation checks (e.g.\ inspection of the blinded observable vector for
obvious systematics) while hiding the cosmological-parameter values which are
eventually to be inferred.

The blinding transformation is described in \eq{eq:addbl}: One simply
adds the difference between two
theoretically-calculated observable vectors to the measured data.
The ``blinding shift'' is the prediction for a reference cosmology \pref subtracted from the prediction for shifted cosmology $\pref+\Delta\params$. In the limit where the measurement noise is invariant and the summary statistics are linear functions of the parameters \params, this transformation will generate blinded summary statistics that are completely indistinguishable from real data generated by the same experiment at an altered set of parameters $\pobs+\Delta\params$, i.e. a perfect blinding. This limit may or may not hold in the region spanned by the reference parameters, the shifted parameters, and the true-sky parameters \pobs.
In practice the model $\datat\left(\params\right)$ for the summary statistics will have non-linearity over the parameter range of interest, and one must check that the blinded data are close (in a \chisq sense) to data that could be generated by \textit{some} valid set of parameters $\Theta_{\rm bl}.$
Therefore, most of this paper is devoted
to verifying that this is the case for the forecasted DES Year 3 galaxy clustering and weak lensing, or \ythree, analysis.
We also check that this blinding transformation is capable of hiding the true results of the analysis, i.e. that this transformation can change the results enough so that that knowing $\Theta_{\rm bl}$ does not allow experimenters to know how the data compare to their prejudices.

These results serve both as a concrete example of the blinding transformation and as verification that it is reliable enough to use for the real DES \ythree analysis. We focus on blinding shifts 
in $\sigma_8$ and $w$, and performed this test by finding the best-fitting parameters for blinded  and unblinded versions of 100 realizations of noiseless synthetic measured observable vectors. (See \fig{fig:chainplot_fid_nice} for an illustrative example.) In order to mimic the fact that the true cosmology will not match our fiducial \pref, we vary the input parameters \pobs for those simulations. For a proof-of-concept fiducial test, we select a subset of cosmologial parameters from a Fisher-forecast-based multivariate Gaussian distribution, and  for a more conservative follow-up ``nuisance test,'' we additionally draw the values of nuisance parameters from flat probability distributions, resulting in very large $\pobs-\pref$ offsets. 
The change in goodness of fit \dchisq due to blinding serves as our principle metric for successful blinding, as it measures the extent to which blinding preserves the internal consistency of the individual observable vector components. 

For the fiducial test,  the impact of blinding on the goodness of
fit was below the expected rms statistical fluctuations in \chisq for all realizations (\fig{fig:dchisqhist_fid}), and the change in best-fitting parameters was generally well predicted by the input blinding shift (\fig{fig:inoutshifts_fid}).  As expected from the leading-order analysis for non-linear models (\sect{sec:lop}), the \dchisq figure of merit for blinding scales as a power of the product of the sizes of the blinding shift $\Delta\params$ and the ``truth shift''  $\pobs-\pref$.
We also verified that those results are not significantly affected when Gaussian noise is added to the simulated data  (\app{app:noisy}).
We noted one potential cause for concern, in that for a small number of realizations with large truth offset $\pobs-\pref$,  the blinding transformation resulted in large changes to the best-fitting values of nominally unblinded parameters which were large enough to push the posterior into a prior boundary.

When we stress the blinding transformation with the nuisance-parameter test case,
the typical \dchisq performance worsened somewhat
(\figs{fig:dchisqhist_nuis}--\ref{fig:dist-dist-dchisq_nuis}).
The majority of realizations in this latter test still fall below the criterion  $\dchisq<30$ (i.e. less than the standard deviation of \chisq), but 4 out of 100 realizations exceeded that value, indicating a poor fit of our model to the blinded data.  These realization are found, however, to have
input parameters \pobs that are very far from the reference parameters \pref assumed for blinding, both in the sense that they would have low probability under the priors on nuisance parameters, and that they are the equivalent of $50\sigma$ unlikely for the \ythree parameter errors at \pref.  An appropriate choice of \pref will preclude the appearance of these high-\dchisq cases in the real \ythree analysis.  

While we have demonstrated that \eq{eq:addbl} defines a viable blinding scheme for \ythree $w$CDM analyses,
it is possible that other experiments will encounter cases where the non-linearities in the data model generate unacceptably large \dchisq over the ranges of $\Delta\params$ and \pobs which are necessary for effective blinding over the full parameter  space allowed by priors. 
 This issue is not necessarily prohibitive, however: one can imagine additional steps in the blinding procedure to account for it (and which could potentially also be used to handle cases where prior-boundary collisions occur even with acceptably small \dchisq). For example, if all pipeline checks pass on blinded data, but the resultant $\Theta_{\rm bl}$ 
 encroaches on the boundaries of priors on nuisance parameters, a designated member of the collaboration could look at how the posterior changes when the data are blinded using a different randomly drawn blinding shift $\Delta \params$.
Alternatively, the person tasked with generating the blinding shifts could generate two or more $f^{\rm (add)}$ shifts
that use \pref values in distinct parts of the parameter space.  The
collaboration's criterion could be that the data are accepted if any one of
these blinded data sets generate an acceptable \chisq.  This would allow valid
data to pass when its \pobs is sufficiently close to any one of the \pref's,
while truly systematic errors would still be likely to fail the \dchisq
criterion.  Other strategies may be viable as well.

There are a number of considerations one should take into account when deciding whether and how to  adopt the summary-statistic blinding transformation described in this paper.
Summary-statistic blinding has the advantage that there is relatively low overhead for implementation:
it can make use of existing analysis infrastructure, since the blinding factors are computed using the same theory prediction machinery needed for parameter estimation. To make the blinding more robust, it should ideally be implemented as an automatic step when summary statistics are measured from the data.  It is a trivial matter for a collaboration member to infer the blinding shifts and unblind their data, if the blinding code is freely available: All one needs to do is run a zero vector through the blinding subroutine.  Thus some level of self-control and trust are still required for successful blinding---we are not proposing a foolproof cryptographic system.

It is also worth considering whether multiple stages of blinding should be adopted: Especially for a new blinding technique and new analyses, having a step of parameter-level blinding (hiding numbers on parameter constraint contours) even after unblinding the summary statistics, can be useful. It is also worth considering how to check or protect against spurious cases where this transformation may lead to undesired behaviour, like pushing an acceptable unblinded parameter values
past its bound from the prior.
It is also important to keep in mind that blinding necessarily adds time to an analysis and in particular that using this transformation  will require MCMC chains for parameter estimation to be re-run when it is time to unblind.  One can argue that this serves as a feature rather than a bug: The barrier to unblinding can help force a collaboration of busy people with divided attention to pause and consider the status of an analysis before proceeding. 

This summary-statistic blinding transformation is implemented as part of the ongoing DES \ythree analysis. In practice, the blinding transformation is applied using a script that runs automatically  when  the 2PCF are measured from galaxy catalogues. This script uses a string seed to pseudo-randomly draw a blinding shift in parameter space, which uses the same configuration files as the parameter estimation pipeline to compute and apply blinding factors to the measured 2PCF. The same transformation will also be applied to the combined analysis of the \mpp data with CMB lensing.
Looking further forward, summary statistic blinding has the potential for broad applicability to many kinds of multiprobe cosmological analyses. It would be interesting to study its applicability to summary statistics for observables beyond 2PCF, such as
supernovae, galaxy cluster number counts, or spectroscopic galaxy clustering measurements.

Blinding to protect results against human bias is essential in modern observational cosmology, where complex analyses combining data from multiple observables are leveraged to make increasingly precise constraints.
Whereas powerful blinding techniques had already been
 devised in experimental particle physics, they do not naturally translate to
cosmology, particularly to multiprobe analyses that can generally not be
separated into distinct ``signal'' and ``background'' domains. The blinding transformation described in this paper provides a new and promising method for blinding such analyses, which we have demonstrated is applicable to current multiprobe analyses like those being done for DES.
An important property of this blinding transformation is
 it becomes more effective (lower \dchisq, see \sect{sec:lop}) as experiments evolve to higher precision and our priors and prejudices focus on smaller regions of parameter space, and non-linear components of the model shrink in comparison to the linear. This makes it a promising potential tool for future cosmological analysis. Of course, one should
explicitly investigate how the performance of summary statistic blinding changes as noise
on the data decreases to levels like one might expect for future surveys
like DESI, LSST, Euclid, and WFIRST.

\section*{Acknowledgments}
This paper has gone through internal review by the DES collaboration.

JM has been supported by the Porat Fellowship at Stanford University and by the Rackham Graduate School at the University of Michigan through a Predoctoral Fellowship. 
GB has been supported by grants AST-1615555 from the US National
Science Foundation, and DE-SC0007901 from the US Department of Energy.
DH has been supported by DOE under Contract DE-FG02-95ER40899 and NSF
under contract AST-1813834.

The analysis and framing of this paper benefited from discussions at the ``Blind Analysis in High Stakes Survey Science: When, Why, and How?'' workshop\footnote{kipac.github.io/Blinding} held in March 2017 at KIPAC/SLAC. We thank the organizers and attendees of that workshop for sharing their insight and experiences.

The analysis made use of the software tools  {\sc SciPy}~\citep{Jones:2001}, {\sc NumPy}~\citep{Oliphant:2006},  {\sc Matplotlib}~\citep{Hunter:2007}, {\sc GetDist}~\citep{Lewis:2019}, {\sc \sc Multinest}~\citep{Feroz:2007kg,Feroz:2008xx,Feroz:2013hea}, {\sc CosmoSIS}~\citep{Zuntz:2014csq}, and {\sc Cosmolike}~\citep{Krause:2016jvl}.
It was supported in part through computational resources and
services provided by Advanced Research Computing at the University of
Michigan, Ann Arbor; the National Energy Research Scientific Computing Center (NERSC), a U.S. Department of Energy Office of Science User Facility operated under Contract No. DE-AC02-05CH11231; and the Sherlock cluster, supported by Stanford University and the Stanford Research Computing Center. We would like to thank all of these facilities for providing computational resources and support that contributed to these research results. 

Funding for the DES Projects has been provided by the U.S. Department of Energy, the U.S. National Science Foundation, the Ministry of Science and Education of Spain, the Science and Technology Facilities Council of the United Kingdom, the National Center for Supercomputing Applications at the University of Illinois at Urbana-Champaign, the Kavli Institute for Cosmological Physics at the University of Chicago, Financiadora de Estudos e Projetos, Fundacao Carlos Chagas Filho de Amparo a Pesquisa do Estado do Rio de Janeiro , Conselho Nacional de Desenvolvimento Cientifico e Tecnologico and the Ministerio da Ciencia e Tecnologia, and the Collaborating Institutions in the Dark Energy Survey.

The Collaborating Institutions are Argonne National Laboratories, the University of Cambridge, Centro de Investigaciones Energeticas, Medioambientales y Tecnologicas-Madrid, the University of Chicago, University College London, DES-Brazil, Fermilab, the University of Edinburgh, the University of Illinois at Urbana-Champaign, the Institut de Ciencies de l'Espai (IEEC/CSIC), the Institut de Fisica d'Altes Energies, the Lawrence Berkeley National Laboratory, the University of Michigan, the National Optical Astronomy Observatory, the Ohio State University, the University of Pennsylvania, the University of Portsmouth, and the University of Sussex.

\appendix
\section{Leading-order behaviour of blinded observable vectors}
\label{app:lo}
We can safely assume that the model $\datavect(\params)$ is
analytic and can be expanded about the reference parameters \pref\ in
a Taylor series.  We can, without loss of generality, set $\pref=0$
and $\datavect(\pref)=0$ in this section.  The Taylor expansion
becomes 
\lneqb\begin{equation}
  \datat_i(\params)  = \datat_{i,\alpha}\theta_\alpha +
                      \frac{1}{2}
                      \datat_{i,\alpha\beta}\theta_\alpha\theta_\beta
                      + O\left(\theta^3\right),
\label{taylor}
\end{equation}\lneqe
where we adopt the usual conventions that repeated indices within a
term indicate summation, and indices after the comma denote
derivatives taken at the reference parameters.
For clarity we will use Latin indices to indicate
dimensions in data space, and Greek indices for parameter-space
dimensions.  We will also, in this section, define
\lneqb\begin{align}
  \vec{s} &= \Delta\params\nn 
  \vec{t} &=  \pobs - \pref, \nonumber
\end{align}\lneqe
the latter being the true cosmology for the observed Universe.  In the
noise-free case, we can apply the quadratic approximation in
(\ref{taylor}) to the blinding \eq{eq:addbl} to obtain the
blinded observable vector
\lneqb\begin{equation}
    \datam^{\rm bl}_i = \datat_{i,\alpha} \left(s_\alpha + t_\alpha\right)
      + \frac{1}{2}\datat,_{i,\alpha\beta} \left( s_\alpha s_\beta +
        t_\alpha t_\beta \right).
\end{equation}\lneqe
We seek the best-fitting parameters $\Theta^{\rm bl}$ by minimizing the
$\chi^2$ of the solution:
\lneqb\begin{equation}
  \chi^2\left(\Theta^{\rm bl}\right)  = \left[ \datam^{\rm bl}_i -  \datat_i \left({\Theta^{\rm
  bl}}\right)\right] F_{ij} \left[ \datam^{\rm bl}_j - \datat_j \left({\Theta^{\rm
                                       bl}}\right)\right], \label{chisqlo}
\end{equation}\lneqe
where we define  $F=\covdat^{-1}$ to be the inverse of the
observational covariance matrix.  We take $F$ to be independent of the
model parameters. In the linear limit it is easy to see that the
blinding shift is always exact, $\Theta^{\rm bl} = \vec{s} + \vec{t},$
so we introduce the correction term $\tilde\theta$ such that
\lneqb\begin{equation}
  \tilde\theta = \Theta^{\rm bl} - \vec{s} - \vec{t}.
\end{equation}\lneqe
With this definition, we can write the data differential to leading
order in each of $\tilde\theta, s,$ and $t$ as
\lneqb\begin{equation}
   \datam^{\rm bl}_i -\datat_i \left({\Theta^{\rm
  bl}}\right) \approx -\datat_{i,\alpha} \tilde\theta_\alpha -
\datat_{i,\alpha\beta}\left[ s_\alpha t_\beta + \tilde\theta_\alpha \left(s_\beta +
    t_\beta\right) + \frac{1}{2} \tilde\theta_\alpha
  \tilde\theta_\beta\right].
\end{equation}\lneqe
Upon substituting this Taylor expansion back into (\ref{chisqlo}), we
can find the blinding shift adjustment $\tilde\theta$ that yields the
minimal $\chi^2$.  Again retaining only leading-order terms in
$\tilde\theta, s,$, and $t$:
\lneqb\begin{align}
  \tilde\theta_i & \approx -\left(D^T F D\right)^{-1} D^T \covdat^{-1} \vec{q}
  \label{tildetheta} \\
    \begin{split}  
  \dchisq & \approx \left(P\vec{q}\right)^T \covdat^{-1} \left(P\vec{q}\right)\\
  & \le \vec{q}^T \covdat^{-1} \vec{q}
                 \label{dchisqmax}
    \end{split}                 
\end{align}\lneqe
where we have defined the first derivative matrix and the quadratic
data perturbation respectively as
\lneqb\begin{align}
  D_{i\alpha} & \equiv \datat_{i,\alpha} \nn
  q_i & \equiv \datat_{i,\alpha\beta}s_\alpha t_\beta. \nonumber
\end{align}\lneqe
$P$ is a projection matrix that removes the portion of the
non-linear data shift $\vec{q}$ which can be fitted by a
shift in parameters:
\lneqb\begin{equation}
  P \equiv I - D \left(D^T F D\right)^{-1} D^T \covdat^{-1}.
  \label{projectorlo}
\end{equation}\lneqe

From these equations, several properties of the additive blinding
transformation are apparent.  First, the transformation is exact, in the sense that
$\dchisq=0,$ in \emph{any} of the following conditions:
\begin{enumerate}
\item The blinding shift  $\vec{s}=\Delta\params$ is zero.
\item The true cosmology equals the reference cosmology,
  $\vec{t}=\pobs-\pref=0.$
\item The model is linear, $\datat_{i,\alpha\beta}=0.$
\item The derivative matrix is invertible, in which case the projector
  matrix $P=0$, because any point in data space can be fit exactly
  with proper choice of parameters, at least locally.
\end{enumerate}

Secondly, we see that a quadratic term in the data model leads to a
deviation $\tilde\theta$ between the naive blinded cosmology
parameters estimate $\Theta^{\rm bl} = \pobs + \Delta\params$ which
scales as the product of the blinding shift $\vec{s}$ and the ``truth
shift'' $\vec{t}.$

Thirdly, the $\dchisq$ in a quadratic approximation to the data
model will scale as the product of the squares of the two shifts, and
inversely with the measurement covariance matrix \covdat,
$\dchisq \propto s^2 t^2 / \covdat.$  The constant of proportionality will depend on the
relations between the directions of the blinding shift, the truth
shift, the curvature of the model, and the covariance matrix of the
observations.  If we include terms beyond quadratic in the data model,
we will find the dependence of \dchisq is of even higher order in $st.$

\section{Modelling for 3x2pt observable vector}\label{app:twoptcalcs}

The theory predictions for the \mpp observable vectors are computed as follows. First, the non-linear matter power  spectrum $P(k,z)$ is computed using a combination of  CAMB~\citep{Lewis:1999bs,Howlett:2012mh} and halofit~\citep{Takahashi:2012em}. Then, using the Limber approximation, we integrate to obtain the angular power spectra between the sets of tracers we are studying. For the correlation between the $i$th redshift bin of tracer $A$ and the $j$th bin of tracer $B$ the angular power spectrum is
\lneqb\begin{equation}
  C_{A B}^{ij}(\ell) = \left.\int \,dz\frac{H(z)}{c\,\chi^2(z)}W^i_{A}(z)W^j_{B}(z)P(k,z)\right|_{k=(\ell+\tfrac{1}{2})/\chi(z)},
\end{equation}\lneqe
where $\chi$ is the comoving radial distance and the weight functions for
galaxy number density $g$ and weak lensing convergence $\kappa$  are defined as
\lneqb\begin{align}
    W^i_g(z,k) &=  n_i(z)\,b_i,\\[0.2cm]
    \begin{split}
  W^i_{\kappa}(z) &= \left(\frac{3H_0^2\Omega_{m}}{2c}\right)
  \left(\frac{\chi(z)}{a(z)\,H(z)}\right) \\
  &\times \int_z^{\infty}dz'n_i(z')\frac{\chi(z') - \chi(z)}{\chi(z')}.\label{lcdm:eq:Wkappa}
  \end{split}
\end{align}  \lneqe
Here $n_i(z)$ is the normalised redshift distribution and $b_i$ is the galaxy bias of galaxies in bin $i$. We then perform Fourier transformations to convert these angular spectra into real-space angular correlation functions, which can be compared to data. The galaxy-galaxy correlation is
\lneqb\begin{equation}
  w^{ij}(\theta) = \sum_{\ell}\frac{2\ell+1}{4\pi}P_{\ell}(\cos{\theta})\,C_{gg}^{ij}(\ell),
  \end{equation}\lneqe
where $P_{\ell}(x)$ is a Legendre polynomial of order $\ell$. In the flat-sky approximation, where sums over spherical harmonics are converted to two-dimensional Fourier modes, the predicted angular correlations  between the shears of galaxies in tomographic bins $i$ and $j$ are 
\lneqb\begin{align}
  \xi_+^{ij}(\theta) &= \int\frac{d\ell\, \ell}{2\pi}J_0(\ell\theta)\, C_{\kappa \kappa}^{ij}(\ell),\\
  \xi_-^{ij}(\theta) &= \int\frac{d\ell\, \ell}{2\pi}J_4(\ell\theta)\, C_{\kappa \kappa}^{ij}(\ell),\\
  \gamma_t^{ij}(\theta) &= \int\frac{d\ell \,\ell}{2\pi}J_2(\ell\theta)\, C_{g \kappa}^{ij}(\ell),
\end{align}\lneqe
where $J_m(x)$ is a Bessel function of the first kind of order $m$. In practice these calculations are done using the function {\tt tpstat\_via\_hankel}: from the nicaea software\footnote{\tt www.cosmostat.org/software/nicaea}~\citep{Kilbinger:2008gk}.

Nuisance parameters are included  as follows. The photo-$z$ bias parameters $\Delta z_i^x$, where $x=$source or lens, have the effect of shifting the redshift distributions of the samples of galaxies:
\lneqb\begin{equation}
  n^x_i(z) \quad\rightarrow\quad n^x_i(z-\Delta z^x_i) 
\end{equation}\lneqe
Shear calibration parameters $m$ are defined so the measured shear for a galaxy is $\gamma_{\rm
   meas}=(1+m)\gamma_{\rm true}$.  They modify the 2PCF involving source galaxies via
 \lneqb\begin{align}
   \xi_{\pm}^{ij}(\theta)&\rightarrow (1+m_i)(1+m_j)\xi_{\pm}^{ij}(\theta), \quad \text{ and }\\
   \gamma_t^{ij}(\theta)&\rightarrow (1+m_j)\gamma_t^{ij}(\theta). 
 \end{align}\lneqe

The linear intrinsic alignment model used in our analysis modifies the shear convergence weight function (\eq{lcdm:eq:Wkappa})  via
 \lneqb\begin{equation}
   W^i_{\kappa}(z) \rightarrow W^i_{\kappa}(z) - \left[A_{IA}\left(\frac{1+z}{1+z_0}\right)^{\alpha_{IA}}\frac{C_1\rho_{m0}}{D(z)}\right]\frac{dn^i}{dz},
 \end{equation}\lneqe
 where $C_1 = 0.0134/\rho_{\rm crit}$ is a normalisation constant calibrated based on previous observations~\citep{Bridle:2007ft}.

 \section{Additional results for noisy data}\label{app:noisy}

 We performed a variation of our fiducial test with noisy measured observable vectors. Here we used the same true cosmology \pobs and blinding shifts $\Delta\params$ as in \sect{sec:results_fid}. The only difference is that after computing the theory prediction at \pobs to generate a synthetic measured observable vector, we added a realization of Gaussian noise produced using a Cholesky decomposition of the  covariance \covdat. 

\begin{figure}
  \centering
\includegraphics[width=\linewidth]{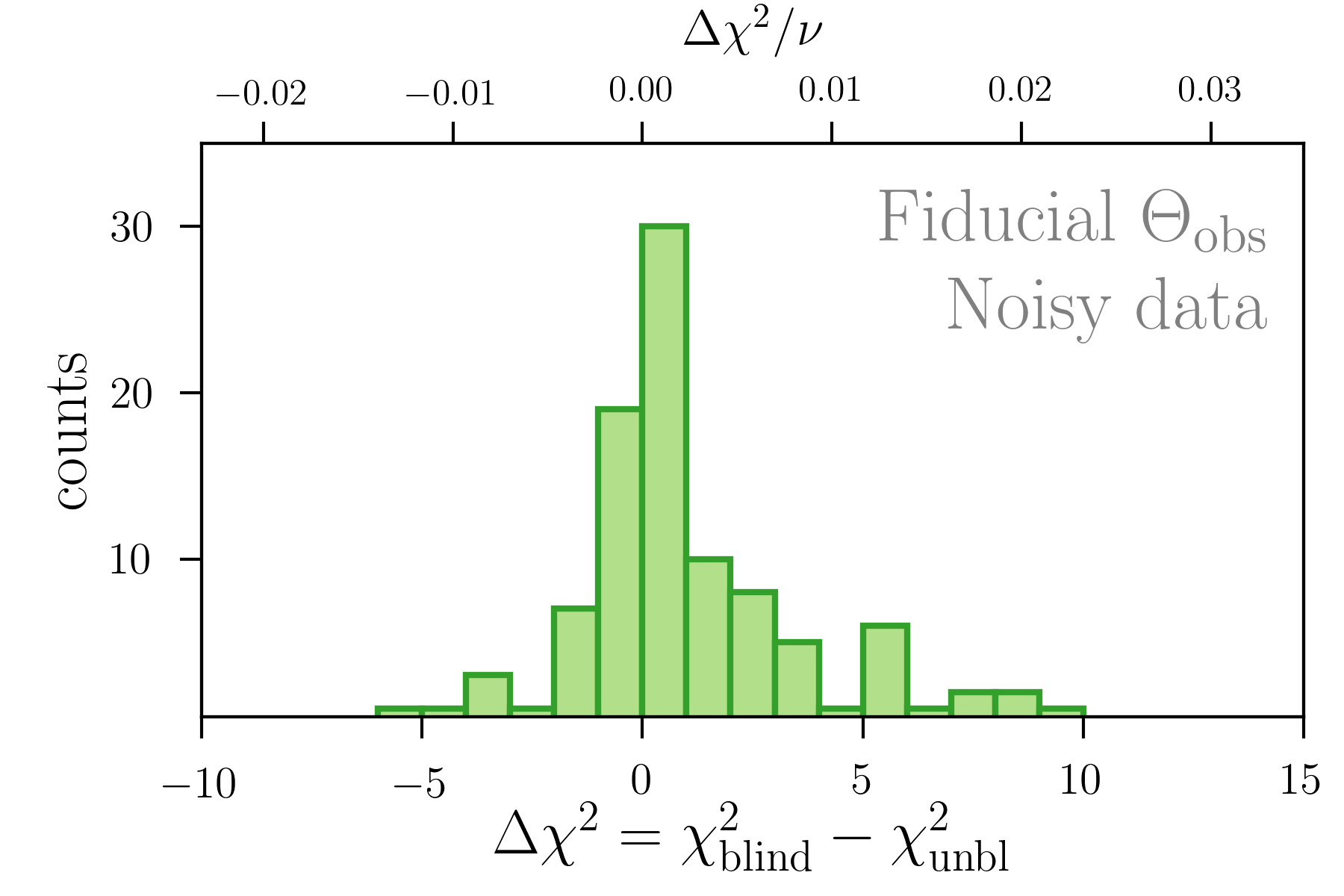}
\caption{Same as \fig{fig:dchisqhist_fid}, but with Gaussian noise added to the fiducial ensemble of unblinded observable vectors: histogram of \dchisq values. }
  \label{fig:dchisqhist_noise}
\end{figure}

\begin{figure}
    \centering
\includegraphics[width=\linewidth]{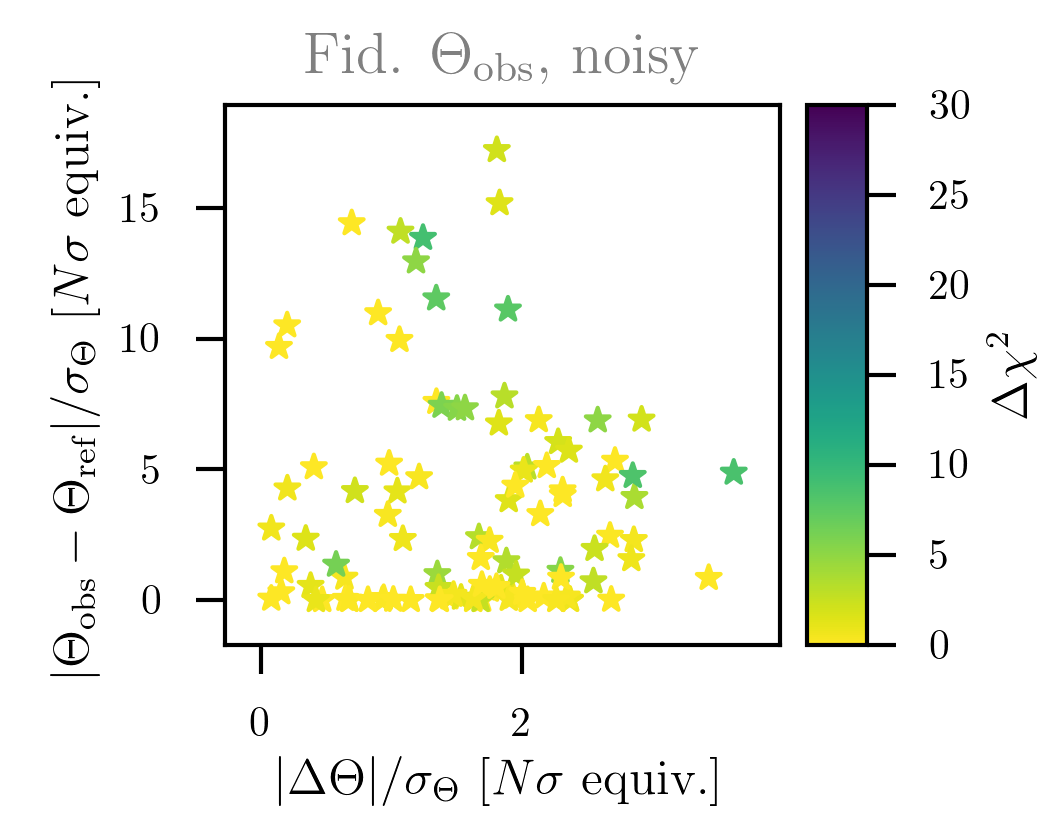}
\caption{Same as \fig{fig:dist-dist-dchisq_fid}, but with Gaussian noise added to the ensemble of unblinded observable vectors:  dependence of \dchisq on the parameter-space distances associated with $\pobs-\pref$ and $\Delta\params$. }
  \label{fig:dist-dist-dchisq_noise}
  \end{figure}

 Results for the noise-added version of our fiducial test are shown here.  \fig{fig:dchisqhist_noise} shows a histogram of the resulting \dchisq values. 
 \fig{fig:dist-dist-dchisq_noise} shows how \dchisq depends on the magnitude of the distances in parameter space associated with the blinding shift $\Delta\params$ and the difference between the true cosmology and that assumed for blinding $\pobs-\pref$. There is more scatter in the relations, but otherwise the results for this test are not substantially different from those shown for the noiseless test presented in \sect{sec:results_fid}. We additionally confirmed for a few realizations that adding noise to the observable vector does not significantly change how blinding affects posterior contours like those shown in \fig{fig:chainplot_fid_nice}.
 
\bibliographystyle{mn2e_plus_arxiv}
\bibliography{literature}
\section*{Affiliations}

$^{1}$ Kavli Institute for Particle Astrophysics \& Cosmology, P. O. Box 2450, Stanford University, Stanford, CA 94305, USA\\
$^{2}$ Department of Physics, University of Michigan, Ann Arbor, MI 48109, USA\\
$^{3}$ Department of Physics and Astronomy, University of Pennsylvania, Philadelphia, PA 19104, USA\\
$^{4}$ Department of Physics \& Astronomy, University College London, Gower Street, London, WC1E 6BT, UK\\
$^{5}$ Max Planck Institute for Astrophysics, Karl-Schwarzschild-Stra. 1. 85748 Garching, Germany\\
$^{6}$ Department of Astronomy/Steward Observatory, University of Arizona, 933 North Cherry Avenue, Tucson, AZ 85721-0065, USA\\
$^{7}$ SLAC National Accelerator Laboratory, Menlo Park, CA 94025, USA\\
$^{8}$ Fermi National Accelerator Laboratory, P. O. Box 500, Batavia, IL 60510, USA\\
$^{9}$ Instituto de Fisica Teorica UAM/CSIC, Universidad Autonoma de Madrid, 28049 Madrid, Spain\\
$^{10}$ LSST, 933 North Cherry Avenue, Tucson, AZ 85721, USA\\
$^{11}$ Physics Department, 2320 Chamberlin Hall, University of Wisconsin-Madison, 1150 University Avenue Madison, WI  53706-1390\\
$^{12}$ CNRS, UMR 7095, Institut d'Astrophysique de Paris, F-75014, Paris, France\\
$^{13}$ Sorbonne Universit\'es, UPMC Univ Paris 06, UMR 7095, Institut d'Astrophysique de Paris, F-75014, Paris, France\\
$^{14}$ Centro de Investigaciones Energ\'eticas, Medioambientales y Tecnol\'ogicas (CIEMAT), Madrid, Spain\\
$^{15}$ Laborat\'orio Interinstitucional de e-Astronomia - LIneA, Rua Gal. Jos\'e Cristino 77, Rio de Janeiro, RJ - 20921-400, Brazil\\
$^{16}$ Department of Astronomy, University of Illinois at Urbana-Champaign, 1002 W. Green Street, Urbana, IL 61801, USA\\
$^{17}$ National Center for Supercomputing Applications, 1205 West Clark St., Urbana, IL 61801, USA\\
$^{18}$ Institut de F\'{\i}sica d'Altes Energies (IFAE), The Barcelona Institute of Science and Technology, Campus UAB, 08193 Bellaterra (Barcelona) Spain\\
$^{19}$ INAF-Osservatorio Astronomico di Trieste, via G. B. Tiepolo 11, I-34143 Trieste, Italy\\
$^{20}$ Institute for Fundamental Physics of the Universe, Via Beirut 2, 34014 Trieste, Italy\\
$^{21}$ Observat\'orio Nacional, Rua Gal. Jos\'e Cristino 77, Rio de Janeiro, RJ - 20921-400, Brazil\\
$^{22}$ Department of Physics, IIT Hyderabad, Kandi, Telangana 502285, India\\
$^{23}$ Excellence Cluster Origins, Boltzmannstr.\ 2, 85748 Garching, Germany\\
$^{24}$ Faculty of Physics, Ludwig-Maximilians-Universit\"at, Scheinerstr. 1, 81679 Munich, Germany\\
$^{25}$ Jet Propulsion Laboratory, California Institute of Technology, 4800 Oak Grove Dr., Pasadena, CA 91109, USA\\
$^{26}$ Santa Cruz Institute for Particle Physics, Santa Cruz, CA 95064, USA\\
$^{27}$ Institut d'Estudis Espacials de Catalunya (IEEC), 08034 Barcelona, Spain\\
$^{28}$ Institute of Space Sciences (ICE, CSIC),  Campus UAB, Carrer de Can Magrans, s/n,  08193 Barcelona, Spain\\
$^{29}$ Kavli Institute for Cosmological Physics, University of Chicago, Chicago, IL 60637, USA\\
$^{30}$ Department of Astronomy, University of Michigan, Ann Arbor, MI 48109, USA\\
$^{31}$ Department of Physics, Stanford University, 382 Via Pueblo Mall, Stanford, CA 94305, USA\\
$^{32}$ Department of Physics, ETH Zurich, Wolfgang-Pauli-Strasse 16, CH-8093 Zurich, Switzerland\\
$^{33}$ Center for Astrophysics $\vert$ Harvard \& Smithsonian, 60 Garden Street, Cambridge, MA 02138, USA\\
$^{34}$ Australian Astronomical Optics, Macquarie University, North Ryde, NSW 2113, Australia\\
$^{35}$ Lowell Observatory, 1400 Mars Hill Rd, Flagstaff, AZ 86001, USA\\
$^{36}$ George P. and Cynthia Woods Mitchell Institute for Fundamental Physics and Astronomy, and Department of Physics and Astronomy, Texas A\&M University, College Station, TX 77843,  USA\\
$^{37}$ Department of Astrophysical Sciences, Princeton University, Peyton Hall, Princeton, NJ 08544, USA\\
$^{38}$ Instituci\'o Catalana de Recerca i Estudis Avan\c{c}ats, E-08010 Barcelona, Spain\\
$^{39}$ Department of Physics and Astronomy, Pevensey Building, University of Sussex, Brighton, BN1 9QH, UK\\
$^{40}$ School of Physics and Astronomy, University of Southampton,  Southampton, SO17 1BJ, UK\\
$^{41}$ Computer Science and Mathematics Division, Oak Ridge National Laboratory, Oak Ridge, TN 37831\\
$^{42}$ Institute of Cosmology and Gravitation, University of Portsmouth, Portsmouth, PO1 3FX, UK\\
$^{43}$ Department of Physics, Duke University Durham, NC 27708, USA\\
$^{44}$ Cerro Tololo Inter-American Observatory, National Optical Astronomy Observatory, Casilla 603, La Serena, Chile\\
$^{45}$ Max Planck Institute for Extraterrestrial Physics, Giessenbachstrasse, 85748 Garching, Germany\\
$^{46}$ Universit\"ats-Sternwarte, Fakult\"at f\"ur Physik, Ludwig-Maximilians Universit\"at M\"unchen, Scheinerstr. 1, 81679 M\"unchen, Germany\\
$^{47}$ Institute for Astronomy, University of Edinburgh, Edinburgh EH9 3HJ, UK\\

\label{lastpage}
\end{document}